\documentclass{jfm-arxiv}
\usepackage{graphicx}
\usepackage{epstopdf,epsfig}
\usepackage[usenames,dvipsnames]{xcolor}
\usepackage{amssymb,amsmath}
\usepackage{siunitx}
\usepackage{placeins}
\usepackage{hyperref}
\hypersetup{hidelinks}

\shorttitle{Sessile drop evaporation in a gap}
\shortauthor{S. Hartmann, C. Diddens, M. Jalaal and U. Thiele}

\title{Sessile drop evaporation in a gap -- crossover between diffusion-limited and phase transition-limited regime}

\author{Simon Hartmann\aff{1,2}
  \corresp{\email{s.hartmann@wwu.de}},
  Christian Diddens\aff{3},
  Maziyar Jalaal\aff{4},
  \and Uwe Thiele\aff{1,2}
  \corresp{\email{uwe.thiele@uni-muenster.de}}}

\affiliation{\aff{1}Institut f\"ur Theoretische Physik, Westf\"alische Wilhelms-Universit\"at M\"unster, Wilhelm-Klemm-Str.\ 9, 48149 M\"unster, Germany
\aff{2}Center of Nonlinear Science (CeNoS), Westf{\"a}lische Wilhelms-Universit\"at M\"unster, Corrensstr.\ 2, 48149 M\"unster, Germany
\aff{3}Physics of Fluids group, Department of Science and Technology, Mesa+ Institute, Max Planck Center for Complex Fluid Dynamics and J. M. Burgers Centre for Fluid Dynamics, University of Twente, P.O. Box 217, 7500 AE Enschede, The Netherlands
\aff{4}{Van der Waals-Zeeman Institute, Institute of Physics, University of Amsterdam Science Park 904, Amsterdam, 1098XH, The Netherlands}}

\begin{document}

\maketitle

\begin{abstract}
  We consider the time evolution of a sessile drop of volatile partially wetting liquid on a rigid solid substrate. Thereby, the drop evaporates under strong confinement, namely, it sits on one of the two parallel plates that form a narrow gap. First, we develop an efficient mesoscopic thin-film description in gradient dynamics form. It couples the diffusive dynamics of the vertically averaged vapour density in the narrow gap to an evolution equation for the profile of the volatile drop. The underlying free energy functional incorporates wetting, interface and bulk energies of the liquid and gas entropy. The model allows us to investigate the transition between diffusion-limited and phase transition-limited evaporation for shallow droplets. Its gradient dynamics character also allows for a full-curvature formulation.
  Second, we compare results obtained with the mesoscopic model to corresponding direct numerical simulations solving the Stokes equation for the drop coupled to the diffusion equation for the vapour as well as to selected experiments. In passing, we discuss the influence of contact line pinning.
\end{abstract}

\section{Introduction}\label{sec:intro}

The dynamics of the liquid-vapour phase change, i.e., evaporation and condensation, plays a very important role in many systems involving films or drops of simple or complex liquids on solid substrates~\citep{Brutin2015}. Examples of practical importance include printing, coating and deposition processes~\citep{BHSF1992jns,Rout2013rpp,Thie2014acis}, as well as cooling, moisture capturing and heat exchange technologies~\citep{OrDB1997rmp,NYPW2018ijts,JaPR2020ijlt}. In consequence, the evaporation of sessile drops of volatile liquids on rigid solid substrates is extensively studied in experiment and theory~\citep{CrMa2009rmp,CaGu2010sm,HuLa2002jpcb,SSVR2011epjt,Erbi2012acis,KoTS2014cocis,Lars2014aj}.

Thereby, the dynamics of droplet evaporation is controlled by the intricate interplay of various transport processes, namely, of heat and material within and between the liquid and the gas phase. They influence interface, temperature and concentration profiles, in turn causing pressure gradients as well as thermal and solutal Marangoni forces~\citep{NepomnyashchyVelardeColinet2002}. These then drive convective motion within the liquid. For droplets on solid substrates, wettability and its interplay with evaporation in the region of the three-phase contact line also plays a crucial role~\citep{POCW2008cec}. Although the involved processes can be modelled employing the full hydrodynamic description based on (Navier-)Stokes equations for the liquid and (advection-)diffusion equations for solutes in the liquid and vapour in the gas phase~\citep{PeBu2008pre,BhFA2009njp}, in many cases reduced descriptions are used. Common examples are long-wave (or lubrication, or thin-film) models for the liquid that are valid for small contact angles and interface slopes~\citep{OrDB1997rmp,CrMa2009rmp,SWVO2016jfm,JiWi2018prf}.

In all cases, the description of the dynamics of evaporating liquid films and drops on solid substrates crucially depends on the model for the evaporation rate \(j_\mathrm{ev}\). It enters the kinematic boundary condition employed at the free liquid-vapour interface~\citep{Levich1962,Leal2007} and gives the mass loss per time and  interface area. The rate \(j_\mathrm{ev}\) depends on material properties, thermodynamic state and on interface and system geometry~\citep{OrDB1997rmp,POCW2008cec,Erbi2012acis}.

One may distinguish two main approaches to the determination of \(j_\mathrm{ev}\) depending on the character of the process that limits the mass transfer across the interface. The limiting step can be either the actual phase change, e.g., for evaporation the transition of molecules from liquid state to gas state, or the diffusive transport of the vapour within the gas surrounding the drop, e.g., for evaporation transport away from the interface~\citep{PiBe1977jcis,SuBA2004jem}. Here, we call the two approaches \textit{(phase) transition-limited} and \textit{diffusion-limited}, respectively. Other important distinctions are (i) whether the process is considered under homogeneously isothermal conditions or whether latent heat and heat transport are incorporated as further rate-limiting influences, and (ii) whether the evaporation is into pure vapour or into an inert gas. Only in the latter case one considers mass diffusion.

Diffusion-limited evaporation is considered in many models for evaporating liquid drops and films on solid substrates, either for simple liquids or suspensions and solutions. Such models are used and analysed, e.g., by \citet{BoSh1995l,DBDH1997n,DBDH2000pre,HuLa2002jpcb,CBPC2002l,PoBC2003l,ErMN2002l,SuBA2004jem,Popo2005pre,HuLa2005l,SRAB2006jfm,MuKo2008pre,EgPi2010pf,SSRA2011csaea,TDRC2015l,SWVO2016jfm}. An overview of earlier work is given by~\citet{HuLa2002jpcb}. This approach assumes that the phase transition is much faster than diffusion, i.e., directly at the liquid-gas interface the vapour is at saturation~\citep{Maxw1890,Lang1918pr}. In consequence, the local evaporation rate along the liquid-gas interface is controlled by the vapour diffusion within the entire gas phase. For shallow macroscopic drops, i.e., in the limit of small contact angles, the evaporation rate has a square-root divergence at the three-phase contact line~\citep{DBDH2000pre,HuLa2002jpcb,Popo2005pre}.\footnote{This corresponds to the small-angle limit of the general expression for the evaporation flux \(j_\mathrm{ev}\sim{(R-r)}^{-\nu}\) with \(\nu=(\upi-2\theta_e)/(2\upi-2\theta_e)\) valid for arbitrary contact angles \(\theta_e\) close to the contact line (see appendix of~\citet{DBDH2000pre} and sec.~8.12 of~\citet{LebedevSilverman2012}). Here $R$ is the drop radius and $r$ the radial coordinate.} Note that, in contrast, drops with contact angle \(\upi/2\) show a uniform rate, as can easily be seen from the electrostatic equivalent using ``mirror charges''. The effect that evaporative cooling has on the saturation concentration and the dependence of the diffusion constant on pressure is incorporated in Refs.~\citep{DWDD2008csaea,SWDD2009pf,DWDS2009pf,DWDD2009jfm}.
Evaporating thin liquid films in a gap geometry are considered by~\cite{SuBA2004jem,SuBB2005jfm} --- in the case of weak surface modulations a nonlocal single thin-film equation of closed form is determined employing Hilbert transforms.
The influence of wettability and capillarity for mesoscale drops was considered by~\citet{EgPi2010pf}. There, a single, though non-local, equation for the dynamics of the thickness profile is obtained.
A comparison of the diffusive and evaporative time scales was discussed by \citet{LeVY2014sm} in terms of a Lattice-Boltzmann model of a volatile drop, yet assuming a diffusion slower than phase change.

The transition-limited case is considered in a number of different flavours: The ``kinetic'' approach by~\citet{BuBD1988jfm,JoDB1991jfm} assumes a uniform constant saturated vapour density in the gas and determines the strength of evaporation/condensation via the difference of film surface temperature and the uniform saturation temperature in the gas phase. The approach is also normally applied if evaporation is into a pure vapour atmosphere, i.e., any vapour dynamics is then neglected. The derivation is based on a discussion of mass, energy and momentum flows across the liquid-vapour interface resulting, e.g., in the incorporation of vapour recoil effects. The approach is adopted in many later works, e.g.,~\citet{AnDa1995pf,Hock1995pf,OrBa1999jcis,WaCM2003jcis,GIMK2006prl,MuKo2008pre,SaRC2017jfm}. Dependencies of evaporation rate on interface curvature and wettability are normally not incorporated.

However, these effects are included in another flavour of the transition-limited case as presented by~\citet{PoWa1972ijhmt,MoHo1980jcis,Wayn1993l,Shar1998l,PaKS1999jcp,KaKS2001l,AjHo2006arfm,JiWi2018prf}. It seems the earliest model for an evaporating meniscus influenced by Laplace (curvature) and Derjaguin (or disjoining) pressures is given by~\citet{PoWa1972ijhmt}, where evaporation into pure vapour is considered.  At the liquid-vapour interface the vapour is at saturation and then varies vertically due to hydrostatic influences, i.e., it is always at equilibrium. The saturation pressure depends on Laplace and Derjaguin pressure. In consequence, Laplace pressure \(\gamma \kappa \) (the product of liquid-gas interface tension $\gamma$ and the interface curvature $\kappa$) and Derjaguin pressure \(\Pi \) enter the evaporation flux \(j_\mathrm{ev}\), however, as argument of an exponential~\citep{Wayn1993l,Shar1998l,PaKS1999jcp,KaKS2001l}. Ultimately, evaporation is driven by a temperature difference between liquid and vapour. So, the process is seen as ``transition-limited'', but what limits the mass transfer between the phases is the diffusion of heat within the liquid. The gas phase itself is always at uniform constant temperature and pressure. Somewhat similar expressions are derived and/or used by~\citet{SaLT1998pre,AjHo2001jcis,LyGP2002pre,Pism2004pre,LeLL2004l,Ajae2005pre,Ajae2005jfm,Thie2010jpcm,ReCo2010mst}, to study dewetting volatile films, vapour bubbles in microchannels and evaporation fronts. There, however, the transition-limitation is indeed due to the phase transition at the interface and the vapour is not at saturation, but at fixed vapour pressure (chemical potential). Furthermore, a direct proportionality of \(j_\mathrm{ev}\) and the sum of \(\gamma \kappa \) and \(\Pi \) is used in the evaporation term.

The majority of works on thin-film models that incorporate evaporation either exclusively consider the transition-limited \textit{or} the diffusion-limited case. A small number of works exist that compare the two approaches~\citep{MuKo2008pre,MuKo2011jfm}.
A partial comparison is also done for a model based on a Stokes description of the liquid~\citep{PeBu2008pre}. The only work we are aware of that develops a more general model containing both limiting cases is by~\citet{SuBB2005jfm}.

As laid out above, most evaporation models assume either (i) limitation by vapour diffusion in the gas phase or (ii) limitation by heat diffusion in the liquid phase, or (iii) limitation by mass transfer between liquid and gas phase. Case (i) is not applicable for evaporation into pure vapour and implicitly assumes uniform total pressure in the gas phase.
Any pressure gradient would trigger convective flows in the gas phase what is, however, excluded. Wetting- and capillarity-influence on saturation concentration can be incorporated. Case (ii) assumes a uniform vapour concentration corresponding to the value at saturation at a gas reference temperature that differs from the temperature of the liquid at the liquid-gas interface. This difference drives evaporation. Wetting- and capillarity-influence on evaporation can be incorporated. Finally, case (iii) assumes constant vapour pressure (chemical potential) in the gas phase what results in inhomogeneous evaporation due to wetting- and capillarity-dependencies. Any evaporation-induced inhomogeneous vapour and total gas pressure are assumed to instantaneously equilibrate. Nearly all mentioned models focus on one of these cases and do not allow for an analysis of transitions between the different limiting cases.

\begin{figure}
  \centering
  \includegraphics[width=0.8\hsize]{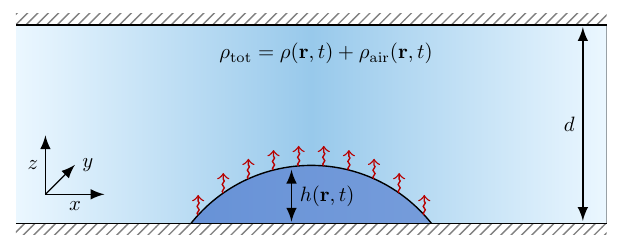}
  \caption{Sketch of the considered system. It consists of two
    parallel rigid smooth solid plates separated by a gap of width \(d\). A layer
    or shallow sessile drop of volatile liquid with a thickness profile \(h(\boldsymbol r,t)\) with $\boldsymbol{r}=(x,y)^T$ is situated on the lower plate without touching the upper one. The height-averaged density between the liquid and the upper plate is \(\rho(\boldsymbol r,t)\) for the vapour, and \(\rho_\mathrm{air}(\boldsymbol r,t)\) for the remaining (inert) gas, i.e., the total height-averaged gas density is \(\rho_\mathrm{tot}=\rho+\rho_\mathrm{air}\). There can be exchange between the liquid and the vapour due to evaporation/condensation.}\label{fig:sketch}
\end{figure}

Our present aim is to develop a relatively simple long-wave model in gradient dynamics form that bridges cases (i) and (iii) for the specific geometry of a sessile drop of partially wetting liquid evaporating into the narrow gap between two parallel rigid smooth solid plates (see Fig.~\ref{fig:sketch}). There, the coupled liquid and gas dynamics can be described by kinetic equations of reduced dimensions. For simplicity, we consider a completely isothermal system -- thermal effects can be incorporated later on.
The simple gradient dynamics approach shall allow one to incorporate our model as a building block into a wide class of thin-film models for more complex settings. Furthermore, we show that the developed simplified mesoscopic approach favourably compares to a full macroscopic description as well as to experiments.

As a macroscopic model we employ a Stokes model coupled to diffusion in the gas phase. Thereby, an evaporation rate $j_\mathrm{evap}$ is implemented into the boundary condition at the liquid-gas interface that is equivalent to the one used in our gradient dynamics model. At the contact line, a Navier slip condition is used. In the gas phase a vapour diffusion model is employed that fully resolves the space within the gap. Versatile variants of this model have been successfully used to account for e.g. multi-component droplet evaporation~\citep{DiddensJFM2017,Li2019} or droplets evaporating on a thin oil film~\citep{Li2020}.

Experimental results on evaporating sessile droplets are extensive (see the reviews of by~\cite{CaGu2010sm,brutin2018recent,zang2019evaporation} ), but only a limited number of the previous studies investigated the effect of confinement inside microfluidic channels~\citep{bansal2017confinement, bansal2017universal, hatte2019lifetime} or in a box~\citep{SRAB2006jfm}. To provide counterpart experiments for our theory, we analyse the evaporation of a droplet between two horizontal plates, where confinement is only imposed in one (vertical) direction.

This paper is structured as follows: In Section~\ref{sec:models} we present the theoretical and experimental approaches that are compared in this work. In particular, Section~\ref{sec:twofield} discusses the general form of long-wave gradient dynamics models for one and two scalar fields with combined conserved and non-conserved dynamics. Then, in Section~\ref{sec:smallgap} we derive the gradient dynamics model for the evaporating drop in the considered small-gap geometry, and specify all necessary parameters and specific functions. The subsequent Sections~\ref{sec:stokes} and~\ref{sec:exp} briefly introduce Stokes description and experimental setup, respectively.
Next, Section~\ref{sec:res-lw} presents results obtained with the developed thin-film model that are in the subsequent Section~\ref{sec:res-stokes-exp} compared with Stokes-equation results and corresponding experiments. Finally, Section~\ref{sec:conc} concludes with a discussion of the limitations of the presented approach and an outlook toward its further development and application.

\section{Models}\label{sec:models}
\subsection{Long-wave gradient dynamics models}\label{sec:twofield}

The dynamics of a layer or shallow drop of nonvolatile liquid in long-wave  approximation~\citep{OrDB1997rmp,CrMa2009rmp} is characterised by the evolution of a single field --- the layer thickness. As first noted by~\citet{OrRo1992jpif} and~\citet{Mitl1993jcis}, the corresponding dynamic equation for the layer thickness \(h\) can be written in gradient dynamics form for a conserved field (see appendix of~\citet{Thie2018csa} for a derivation of this form via Onsager's variational principle \citep{Doi2011jpcm}). To account for evaporation in the phase transition-limited case~\citep{LyGP2002pre}, one adds thermodynamically consistent nonconserved contributions to the dynamics~\citep{Thie2010jpcm} and obtains
\begin{equation}
  \partial_t h \,=\,
  \bnabla\bcdot\left(Q\,\nabla\frac{\delta\mathcal{F}}{\delta h}\right)
  - M\left(\frac{\delta \mathcal{F}}{\delta h} -p_\mathrm{vap}\right),
  \label{eq:onefield:gov}
\end{equation}
where the energy functional \(\mathcal{F}[h]\) contains wetting energy and surface energy of the free liquid-gas interface.  In the simplest case, \(p_\mathrm{vap}\) is the imposed constant external vapour pressure in the gas phase (also the total gas pressure is then constant if evaporation is not into pure vapour).  Here and in the following \(\partial_t\) denotes the partial time derivative and \(\nabla = (\partial_x, \partial_y)^T\) is the two-dimensional (2d) spatial gradient operator. The functions \(Q(h)\ge0\) and \(M(h)\ge0\) are the positive mobilities of the conserved and the non-conserved contributions, respectively. For a discussion of different forms of \(M(h)\) see~\citet{Thie2014acis}.

For systems with more degrees of freedom, the described one-field model~\eqref{eq:onefield:gov} is extended by incorporating the dynamics of further fields \citep{Thie2018csa}. In the context of thin-film hydrodynamics, two-field gradient dynamics models are presented and analysed for (i)  dewetting two-layer films on solid substrates, i.e., staggered layers of two immiscible fluids~\citep{PBMT2004pre,PBMT2005jcp,JHKP2013sjam,BCJP2013epje}, (ii) decomposing and dewetting films of a binary liquid mixture (with non-surface active components)~\citep{Thie2011epjst,ThTL2013prl,DGGR2021l}, (iii) the dynamics of a liquid film that is covered by an insoluble surfactant~\citep{ThAP2012pf,ThAP2016prf}, (iv) the spreading of a liquid drop on a polymer brush~\citep{ThHa2020epjt} and (v) on an elastic substrate without~\citep{HeST2021sm} and with~\citep{HEHZ2022preprint} Shuttleworth effect. In all these cases, the model is of the form
\begin{equation}
  \partial_t u_a = \bnabla \bcdot \left( \sum_{b=1}^2 Q_{ab}
  \nabla \frac{\delta \mathcal{F}}{\delta u_b}\right)
  - \sum_{b=1}^2 M_{ab} \frac{\delta \mathcal{F}}{\delta u_b}
  \label{eq:nn3}
\end{equation}
where the indices \(a, b = 1,2\) refer to the two fields.
For the considered relaxational dynamics, the \(\mathsfbi{Q}(u_1,u_2)\) and \(\mathsfbi{M}(u_1,u_2)\) represent \(2 \times 2\) \textit{positive definite} and \textit{symmetric} mobility matrices for the conserved and non-conserved parts of the dynamics, respectively, written here in terms of their components $Q_{ab}$ and  $M_{ab}$.
In examples (i) to (iii), both fields show a conserved dynamics, i.e., \(\mathsfbi{M}=0\). However, in general, they can also show a purely nonconserved dynamics, i.e., \(\mathsfbi{Q}=0\) (corresponding to a two-field model-A in the \citet{HoHa1977rmp} classification), or a mixed dynamics as in cases (iv) and (v), i.e., \(\mathsfbi{M}, \mathsfbi{Q}\neq0\).

The mobilities \(\mathsfbi{Q}\) enter the fluxes \(\boldsymbol j_a=-\sum_{b=1}^2Q_{ab}\nabla({\delta\mathcal{F}}/{\delta u_b})\) of the conserved part of the dynamics for both fields \(u_a\). They are given as linear combinations of the influences of both thermodynamic forces \(-\nabla({\delta \mathcal{F}}/{\delta u_b})\). The components of \(\mathsfbi{M}\) give the transition rates between the two fields and between the fields and the surroundings. The conserved fields \(u_1\) and \(u_2\) represent in case (i) the lower layer thickness \(h_1\) and overall thickness \(h_2\), respectively~\citep{PBMT2004pre,PBMT2005jcp,JHKP2013sjam} or the lower and upper layer thickness~\citep{BCJP2013epje}. In case (ii), \(u_1\) and \(u_2\) represent the film height \(h\) and the effective solute height \(\psi=ch\), respectively, where \(c\) is the height-averaged solute concentration, while in case (iii), \(u_1\) and \(u_2\) represent the film height \(h\) and the surfactant coverage.respectively~\citep{ThAP2012pf}. Finally, in cases (iv) and (v) $u_1$ represents the drop height while $u_2$ stands for the local amount of liquid in the polymer brush \citep{ThHa2020epjt} and the elastic-liquid interface profile \citep{HeST2021sm}, respectively.

\subsection{Gradient dynamics for volatile liquid in small-gap geometry}\label{sec:smallgap}
\subsubsection{Gradient dynamics form}
Having set the stage for formulating thin-film models as gradient dynamics on an underlying energy functional, we next introduce such a thin-film model for an evaporating sessile liquid drop with profile \(h(\boldsymbol r,t)\) in a gap of width \(d\) (see Fig.~\ref{fig:sketch}). We employ two fields, namely, on the one hand, the amount \(\psi_1(\boldsymbol r,t)\) of the substance in liquid state in the drop per substrate area and, on the other hand, the amount \(\psi_2(\boldsymbol r,t)\) of the substance in vapour form in the gas phase also per substrate area. The field \(\psi_1(\boldsymbol r,t)\) is proportional to the thickness of the liquid film:
\begin{equation}
  \psi_1(\boldsymbol r,t)= \rho_\mathrm{liq} h(\boldsymbol r,t)\label{eq:def_psi_1}
\end{equation}
where \(\rho_\mathrm{liq}\) is the constant liquid density (to be specified later). All employed densities are number densities, i.e., are given in units of particles per volume. The field \(\psi_2(\boldsymbol r,t)\) is proportional to the height-averaged vapour density \(\rho(\boldsymbol r,t)\), namely,
\begin{equation}
  \psi_2(\boldsymbol r,t)= \rho(\boldsymbol r,t) [d-h(\boldsymbol r,t)].
\end{equation}
The gas phase can either consist of pure vapour or of vapour in an inert gas (called ``air'' in the following). The latter has height-averaged density \(\rho_\mathrm{air}(\boldsymbol r,t)\) and the resulting total gas density is \(\rho_\mathrm{tot}=\rho+\rho_\mathrm{air}\). The literature distinguishes three main cases:
\begin{enumerate}
  \item Evaporation into pure vapour treated isothermally. In this case no diffusion occurs, all dynamics in the gas phase is due to pressure equilibration via convective motion. However, this is a very fast process as it occurs with the speed of sound~\citep{Maxw1890}. On the time scale of evaporation one then assumes uniform vapour, i.e., gas pressure. This implies  \(\rho(\boldsymbol r,t)\) is uniform.\label{case1}
  \item Evaporation into air treated isothermally. In this case vapour diffusion is very important. As in case~\eqref{case1}, the total pressure equilibrates fast, i.e.,  also \(\rho_\mathrm{tot}\) is constant and uniform, e.g., \(p=k_B T \rho_\mathrm{tot}\) for an ideal gas. In consequence,
        \begin{equation}
          \rho_\mathrm{air}(\boldsymbol r,t)=\rho_\mathrm{tot}-\rho(\boldsymbol r,t) =
          \rho_\mathrm{tot}-\frac{\psi_2(\boldsymbol r,t)}{d-h(\boldsymbol r,t)}
        \end{equation}\label{case2}
  \item As case~(\ref{case1}) or case~(\ref{case2}), but not treated isothermally. Then, heat diffusion becomes a limiting factor as the heat used up as latent heat during the liquid-vapour phase transition needs to be transported to the interface. It also becomes important, that normally a jump in the heat flux is considered at the interface that ultimately controls the evaporation flux. Here, we will not discuss this case.\label{case3}
\end{enumerate}
Next, we write the model for the coupled dynamics of the local amounts of liquid and vapour in the gradient dynamics form~\eqref{eq:nn3}. In particular, the \(\psi_i\) follow the mixed conserved and nonconserved dynamics
\begin{align}
   & \partial_t \psi_1 =
  \bnabla\bcdot\left(Q_{11}\,\nabla\frac{\delta F}{\delta
    \psi_1}\,+\,Q_{12}\nabla\frac{\delta F}{\delta \psi_2}\right)
  - M_{11}\frac{\delta F}{\delta \psi_1} - M_{12}\frac{\delta F}{\delta \psi_2},
  \nonumber              \\
   & \partial_t \psi_2 =
  \bnabla\bcdot\left(Q_{21}\nabla\frac{\delta F}{\delta
    \psi_1}\,+\,Q_{22}\nabla\frac{\delta  F}{\delta \psi_2}\right)
  -M_{21}\frac{\delta F}{\delta
    \psi_1}\,-\,M_{22}\frac{\delta  F}{\delta \psi_2}.
  \label{eq:twofield-evolution}
\end{align}
Note, however, that \(h\) is the relevant field for many of the physical
effects and that often it may be more convenient to write the governing
equations in terms of \(h\) [Eq.~\eqref{eq:def_psi_1}] and \(\psi=\psi_2 \).

First, we focus on case~(\ref{case2}) before discussing the amendments needed for case~(\ref{case1}).
As on the considered time scales, the total gas pressure is uniform, there is no pressure gradient that would drive convective transport in the gas phase.
In consequence, there is no dynamic coupling between liquid and gas layer, i.e., \(Q_{12}=Q_{2 1}=0\). Transport in the gas phase is then limited to vapour diffusion within the air.
As a result, the conserved mobilities are\footnote{We obtain \(Q_{\psi_1\psi_1}\) by considering
  \(\rho_l\bnabla\bcdot\left(\frac{h^3}{3 \eta}\,\nabla\frac{\delta F}{\delta h}\right) =\rho_l\,\partial_t h = \partial_t \psi_1 = \bnabla\bcdot\left(Q_{\psi_1\psi_1}\,\nabla\frac{\delta F}{\delta \psi_1}\right)\). The form of the diffusive mobility is for solutions discussed by~\citep{Thie2011epjst,XuTQ2015jpcm}. The form $\widetilde D \psi_2=\widetilde D (d-h)\rho$ is the direct equivalent in the present case.}
\begin{equation}
  \boldsymbol{Q}
  =\left(\! \! \!
  \begin{array}{cc}
      \frac{1}{\rho_\mathrm{liq}}\,\frac{\psi_1^3}{3 \eta} & 0                   \\[.3ex]
      0                                                    & \widetilde D \psi_2
    \end{array}
  \! \! \!\right)\label{eq:twofield-mob-c}
\end{equation}
where \(\eta \) is the dynamic viscosity of the liquid and \(\widetilde D\) is a diffusive mobility constant. The latter is related to the usual diffusion constant \(D\) of the vapour particles in air by \(D =  k_B T \widetilde D\) where \(k_B\) is Boltzmann's constant and \(T\) the temperature. In the simplest case, the nonconserved mobilities are
\begin{equation}
  \boldsymbol{M}
  = \widetilde M\left(\! \! \!
  \begin{array}{cc}
      1  & -1 \\[.3ex]
      -1 & 1
    \end{array}
  \! \! \!\right)\label{eq:twofield-mob-nc}
\end{equation}
where \(\widetilde M\) is an evaporation rate constant, which can be estimated e.g. from the Hertz-Knudsen equation~\citep{knudsen1915maximale, librovich2017non}. The particular matrix ensures that phase change is driven by differences in chemical potentials and that \(\psi_1+\psi_2 \) is conserved, i.e., their sum satisfies a continuity equation \(\partial_t (\psi_1+\psi_2)= \bnabla\bcdot(\boldsymbol{j}_{\psi_1}+\boldsymbol{j}_{\psi_2})\).

The free energy in long-wave approximation is
\begin{equation}
  F = \int_\Omega \left[ \underbrace{\frac{\gamma}{2} {(\nabla
      h )}^2}_{\mathrm{surface}} + \underbrace{g(h)}_{\text{wetting}}
+\underbrace{h f_\mathrm{liq}(\rho_\mathrm{liq})}_{\text{liquid bulk}}
  +\underbrace{(d-h) f_\mathrm{vap}(\rho)}_{\text{vapour bulk}}
  +\underbrace{(d-h) f_\mathrm{air}(\rho_\mathrm{air})}_{\text{air bulk}}
  \right] \mathrm{d}^2 r, \label{eq:vapourGap-energy} \end{equation}
where the \(f_i\)'s are bulk liquid, vapour and air energies per volume, \(\gamma \) is the liquid-gas interface tension and \(g(h)\) is the wetting energy per area. The functional is accompanied by the constraint of particle number conservation across the two phases. The condition is
\begin{equation}
  0 = \int_\Omega \left[\psi_1 + \psi_2 - \bar{n}\right]\mathrm{d}\boldsymbol{x}
  = \int_\Omega \left[h \rho_\mathrm{liq} + (d-h) \rho -\bar{n}\right] \mathrm{d}^2 r. \label{eq:vapourGap-mass}
\end{equation}
where \(\bar{n}\) is the mean (liquid and vapour) particle number per substrate area. Here, particle flux through the boundaries of the domain \(\Omega \) is excluded, however, it can be easily incorporated. Also, gravity is neglected as we consider small droplets, but may be added in the form of potential energy.

The presented long-wave formulation of the mesoscopic model is best suited for shallow droplets. However, using a full-curvature trick similar to \citep{GaRa1988ces,SADF2007jfm} one may also study the evaporation of droplets with larger contact angles. Then, the surface energy term in Eq.~\eqref{eq:vapourGap-energy} is written with the full metric factor $\gamma\sqrt{1+(\nabla h )^2}$ and the evaporation rate per surface area $\widetilde{M}$ in Eq.~\eqref{eq:twofield-mob-nc} is replaced by $\widetilde{M}\sqrt{1+(\nabla h )^2}$. Below we will refer to this amendment as the full-curvature formulation of the mesoscopic model.

Further we remark that Eq.~\eqref{eq:vapourGap-energy} is in a somewhat mixed form as it considers at the same time equations of state encoded in the \(f\)'s for liquid and vapour that may result in phase change, but also uses the film height \(h\) as it is the most important quantity for mesoscopic hydrodynamics. As stated above, we treat \(\rho_\mathrm{liq}\) as constant and will consider the vapour as ideal gas. This is the result of a two-step procedure, explained in the next section (that may be skipped by the reader who is mainly interested in the resulting thin-film equations).

\subsubsection{From real to ideal vapour}
\label{sec:tf-real}
An equation of state for a real gas, e.g., a van der Waals gas, predicts coexisting liquid and gas densities at equilibrium. In an out of equilibrium hydrodynamic description, these densities may then vary in space and time. However, as there is no simple way to express the three fields \(\rho_\mathrm{liq}(\boldsymbol r,t), \rho (\boldsymbol r,t)\) and \(h(\boldsymbol r,t)\) in terms of the two fields \(\psi_1 (\boldsymbol r,t)\) and \(\psi_2 (\boldsymbol r,t)\) we follow a two-step procedure:

(i) We assume a thick homogeneous sharp-interface flat film of thickness \(h\) and calculate \(\rho_\mathrm{liq}\) and \(\rho_\mathrm{vap}\) at coexistence (as functions of temperature \(T\)) using
\begin{equation}
  F_0[\rho_\mathrm{liq}, \rho_\mathrm{vap}] = \int_\Omega \left[ h f_\mathrm{liq}(\rho_\mathrm{liq})
    +(d-h) f_\mathrm{vap}(\rho_\mathrm{vap}) \right] \mathrm{d}^2r, \label{eq:vapourGap-energy0}
\end{equation}
Using the trick to treat \(\xi=d-h\) as independent field, we minimise
\begin{align}
  F_1[\rho_\mathrm{liq}, \rho_\mathrm{vap},h,\xi] = & \int_\Omega \left[ h f_\mathrm{liq}(\rho_\mathrm{liq})
    +\xi f_\mathrm{vap}(\rho_\mathrm{vap}) \right] \mathrm{d}^2r\nonumber                                                                        \\
                                                    & - \tilde \mu \int_\Omega \left[h \rho_\mathrm{liq} + \xi \rho -\bar{n}\right]\mathrm{d}^2r
  + \tilde p \int_\Omega \left[h + \xi - d\right]\mathrm{d}^2r, \label{eq:vapourGap-energy1}
\end{align}
with respect to \(\rho_\mathrm{liq}, \rho_\mathrm{vap},h\) and \(\xi \). Here, \( \tilde \mu \) and \(\tilde p\) are Lagrange multipliers for mass and volume conservation. We obtain
\begin{equation}
  \begin{aligned}
    \tilde \mu & = f'_\mathrm{liq}(\rho_\mathrm{liq})  = f'_\mathrm{vap}(\rho_\mathrm{vap})                                                                \\
    \tilde p   & = f_\mathrm{liq}(\rho_\mathrm{liq}) -  \tilde \mu\rho_\mathrm{liq}   = f_\mathrm{vap}(\rho_\mathrm{vap}) -  \tilde \mu \rho_\mathrm{vap},
  \end{aligned}
\end{equation}
i.e., the standard Maxwell construction for phase coexistence.

In this step we can either employ a single function \(f=f_\mathrm{vap}=f_\mathrm{liq}\) that allows for a liquid-gas phase transition (e.g.\ the van der Waals free energy) or use a purely entropic \(f_\mathrm{vap}(\rho) = k_B T \rho[\log(\Lambda^3 \rho)-1]\) and combine it with a \(f_\mathrm{liq}\) that allows for a phase transition. These we use to obtain the coexisting \(\rho_\mathrm{liq}\) and \(\rho_\mathrm{vap}\) analytically or (most likely) numerically.  From hereon, symbols \(\rho_\mathrm{liq}\), \(\rho_\mathrm{vap}\) and \(f_\mathrm{liq}\) denote the values at coexistence.

(ii) Next we approximate the equation of state in the vapour and the liquid phase. For the liquid phase we neglect any compressibility, i.e., we fix the density in the liquid layer to \(\rho_\mathrm{liq}\). With other words the ``liquid branch'' of the equation of state \(p(\rho)\) is replaced by a vertical line at \(\rho=\rho_\mathrm{liq}\).  The ``gas branch'' of the equation of state is either directly replaced by the ideal gas law given in the previous paragraph or, alternatively, is expanded up to linear order about \(\rho=\rho_\mathrm{vap}\). This results in a shifted and scaled ideal gas law. The latter approach has the advantage that coexistence pressure and concentration are exactly as in the equation of state one started with.

In this way the relation \(h=\psi_1/\rho_\mathrm{liq}\) introduced above becomes meaningful and the relation of variations w.r.t.~\(h\) and \(\psi_1 \) alluded to earlier is justified. Also using \(\rho=\psi_2/(d-h)\), the free energy functional Eq.~\eqref{eq:vapourGap-energy} is written as
\begin{align}
  F[\psi_1,\psi_2] = & \int_\Omega \Bigg[ \frac{\gamma}{2 \rho_\mathrm{liq}^2} {(\nabla
      \psi_1)}^2 + g\left(\frac{\psi_1}{\rho_\mathrm{liq}}\right)
    +\frac{f_\mathrm{liq}}{\rho_\mathrm{liq}} \psi_1
    +\left(d-\frac{\psi_1}{\rho_\mathrm{liq}}\right)
    f_\mathrm{vap}\left(\frac{\psi_2}{d-\psi_1/\rho_\mathrm{liq}}\right) \nonumber                                                                             \\
                     & +\left(d-\frac{\psi_1}{\rho_\mathrm{liq}}\right) f_\mathrm{air}\left(\rho_\mathrm{tot}-\frac{\psi_2}{d-\psi_1/\rho_\mathrm{liq}}\right)
    \Bigg] \mathrm{d}^2 r. \label{eq:vapourGap-energy-psi}
\end{align}
Here, it only depends on \(\psi_1 \) and \(\psi_2 \). Note that \(f_\mathrm{liq}\) is now a constant given by the Maxwell construction. Next we bring all the information together and present the thin-film equations.

\subsubsection{Thin-film equations}
The energy functional~\eqref{eq:vapourGap-energy-psi} in terms of \(\psi_1 \) and \(\psi_2 \) is now minimised together with the particle number constraint~\eqref{eq:vapourGap-mass} (Lagrange multiplier \(\mu \))
w.r.t.~variations in the two fields. This gives
\begin{align}
  \frac{\delta F}{\delta \psi_1} = \frac{1}{\rho_\mathrm{liq}} \big[
   & -\gamma\Delta h + g'\left(h\right)
    +f_\mathrm{liq} -f_\mathrm{vap}(\rho)
    + \rho f_\mathrm{vap}'(\rho)\nonumber     \\
   & - f_\mathrm{air}(\rho_\mathrm{tot}-\rho)
    - \rho f_\mathrm{air}'(\rho_\mathrm{tot}-\rho)
    \big] - \mu
  \label{eq:var1}
\end{align}
where the brackets contain Laplace pressure, Derjaguin pressure, liquid energy and vapour pressure. Note that the somewhat unusual final contribution to the vapour pressure is a direct consequence of the assumption of constant \(p_\mathrm{tot}\).
Here and in the following, we use \(h\) and \(\rho\) as abbreviations where appropriate.

The variation w.r.t.~\(\psi_2 \) gives
\begin{equation}
  \frac{\delta F}{\delta \psi_2} =f_\mathrm{vap}'(\rho) - f_\mathrm{air}' (\rho_\mathrm{tot}-\rho) - \mu,
  \label{eq:var2}
\end{equation}
i.e., a difference in chemical potentials. If we now specify vapour and air to be ideal gases,
\(f_\mathrm{vap}=k_B T \rho[\log(\Lambda^3 \rho)-1]\) and \(f_\mathrm{air}=k_B T \rho_\mathrm{air}[\log(\Lambda^3 \rho_\mathrm{air})-1]\) with the mean free path length $\Lambda $, we obtain
\begin{equation}
  \frac{\delta F}{\delta \psi_1} = \frac{1}{\rho_\mathrm{liq}}
  \left\{-\gamma\Delta h+
  g'(h) +f_\mathrm{liq} +  k_B T\rho_\mathrm{tot} - k_B T\rho_\mathrm{tot} \log[\Lambda^3 (\rho_\mathrm{tot}-\rho)]\right\} - \mu.
  \label{eq:var1id}
\end{equation}
The gradient of this pressure drives the dynamics in the liquid film. The constant liquid energy density \(f_\mathrm{liq}\) and the constant ideal pressure \(k_BT\rho_\mathrm{tot}\) do not contribute to the conserved dynamics, but the final term in the curly parenthesis does. It is a direct consequence of treating case~(\ref{case2}), i.e., of imposing a uniform gas pressure, i.e., constant \(\rho_\mathrm{tot}\). The other variation is then
\begin{equation}
  \frac{\delta F}{\delta \psi_2} =k_B T \{\log(\Lambda^3 \rho) - \log[\Lambda^3 (\rho_\mathrm{tot}-\rho)]\} - \mu.
  \label{eq:var2id}
\end{equation}
Also here the second term in the curly parenthesis is a consequence of the imposed uniform pressure in case~(\ref{case2}).

Introducing the obtained expressions into the general two-field gradient dynamics~\eqref{eq:twofield-evolution} with~\eqref{eq:twofield-mob-c} and~\eqref{eq:twofield-mob-nc} gives
\begin{align}
  \partial_t \psi_1 \, = &
  \bnabla\bcdot\left(\frac{\psi_1^3}{3 \rho_\mathrm{liq}^2\eta}\,\nabla \left\{
  -\gamma \Delta h + g'\left(h\right)
  -  k_B T\rho_\mathrm{tot} \log[\Lambda^3(\rho_\mathrm{tot}-\rho)]  \right\}\right)
  - \widetilde M E
  \nonumber                                                                                                                                                                                                        \\
  \partial_t \psi_2 =    &
  \bnabla\bcdot\left(\widetilde D k_B T \frac{d-h}{1-\rho/\rho_\mathrm{tot}}\nabla\rho\right)
  + \widetilde M E
  \label{eq:film-vapour-evolution-case2}                                                                                                                                                                           \\
  \mathrm{where}\, \,
  E=                     & \frac{1}{\rho_\mathrm{liq}} \left[-\gamma \Delta h + g'(h)+f_\mathrm{liq}\right] \nonumber                                                                                              \\
                         & +                       k_B T  \left\{ \frac{\rho_\mathrm{tot}}{\rho_\mathrm{liq}} +\left(1- \frac{\rho_\mathrm{tot}}{\rho_\mathrm{liq}}\right) \log[\Lambda^3(\rho_\mathrm{tot}-\rho)]
  - \log(\Lambda^3 \rho) \right\}
  \nonumber
\end{align}
This is the final result for case~(\ref{case2}).

Then, the saturation vapour density \(\rho_\mathrm{sat}\) above a flat thick film is obtained by setting the transfer term \(E=0\) and dropping capillarity and wettability influences:
\begin{equation}
  E = \frac{f_\mathrm{liq}}{\rho_\mathrm{liq}}
  + k_B T \left\{ \frac{\rho_\mathrm{tot}}{\rho_\mathrm{liq}} +\left(1- \frac{\rho_\mathrm{tot}}{\rho_\mathrm{liq}}\right) \log[\Lambda^3(\rho_\mathrm{tot}-\rho_\mathrm{sat})]
  - \log(\Lambda^3 \rho_\mathrm{sat}) \right\}=0.
\end{equation}
The corresponding film height \(H\) is determined by the conservation of mass $\bar n  = \psi_1+\psi_2 = H \rho_\mathrm{liq} + (d-H)\rho_\mathrm{sat}$.

The mentioned somewhat unexpected terms in~\eqref{eq:film-vapour-evolution-case2} are well behaved: The additional contributions in the equation for \(\psi_2 \) provide a factor \(1/[1-\rho/\rho_\mathrm{tot}]\) to the generalised diffusion constant. Normally, \(\rho/\rho_\mathrm{tot}\ll1\) and can be neglected. If, however, \(\rho/\rho_\mathrm{tot}\to1\) we approach the limit of pure vapour where diffusion is not the proper transport process any more: the factor in question diverges, i.e., \(\rho \) becomes instantaneously uniform. The additional term in the equation for \(\psi_1 \) corresponds to a flux proportional to \(\frac{1}{1-\rho/\rho_\mathrm{tot}}\nabla\rho \). For \(\rho/\rho_\mathrm{tot} \ll1\) this is the gradient in partial vapour pressure that drives some flow in the adjacent liquid layer, one can see it as an ``osmotic coupling''. It is normally very small as compared to the other pressure gradients. The local evaporation rate \(\widetilde M E\) contains additional terms proportional to the ratio of total gas density and liquid density \(\rho_\mathrm{tot}/\rho_\mathrm{liq}\) that we expect to be small. For dry air and water the ratio is about \(10^{-3}\), and humid air has an even lower density than dry air.

For \(\rho/\rho_\mathrm{tot}\ll1\) and \(\rho_\mathrm{tot}/\rho_\mathrm{liq}\ll1\), Eqs.~\eqref{eq:film-vapour-evolution-case2} written in terms of \(h\) and \(\rho \) reduce to
\begin{align}
  \partial_t h \,       & =
  -\bnabla\bcdot\left\{\frac{h^3}{3 \eta}\,\nabla \left[
    \gamma \Delta h - g'\left(h\right) \right]\right\}
  - j_\mathrm{evap}
  \nonumber                                                         \\
  \partial_t[(d-h)\rho] & =
  \bnabla\bcdot\left[D (d-h) \nabla\rho\right]
  + \rho_\text{liq} j_\mathrm{evap} \label{eq:film-vapour-evolution-case2simple}
  \\
  \mathrm{where}\, \,
  j_\mathrm{evap}/M     & = -\gamma \Delta h + g'(h)+f_\mathrm{liq}
  - \rho_\mathrm{liq} k_B T \log\left(\frac{\rho}{\rho_\mathrm{tot}-\rho}\right).
  \nonumber
\end{align}
Here, we have introduced the diffusion constant $D = \widetilde D k_B T$ and the evaporation rate constant of $M = \widetilde M / \rho_\text{liq}^2$, thus converting the local particle evaporation rate $\widetilde M E$ into a volume rate $j_\mathrm{evap} = \widetilde M E / \rho_\mathrm{liq}$.
Equations~\eqref{eq:film-vapour-evolution-case2simple} allow one to study the crossover between transition-limited drop evaporation dynamics (small \(\rho \)) and diffusion-limited evaporation dynamics (\(\rho \) at drop close to saturation \(\rho_\mathrm{sat}\)).

In case~(\ref{case2}) considered up to here, convective motion in the gas layer is neglected assuming a uniform total gas pressure \(p_\mathrm{tot}\), i.e., a uniform total gas density \(\rho_\mathrm{tot}=\rho_\mathrm{air}+\rho \) that is an externally controlled parameter of the system. However, this directly couples \(\rho_\mathrm{air}\) to the vapour density \(\rho \) what itself results in small additional contributions to the pressure in the liquid, to the evaporation/condensation rate and vapour diffusion. All these terms are direct consequences of the gradient dynamics structure. Although one may neglect them due to their smallness one needs to keep in mind that this breaks the thermodynamic consistency and, therefore, can result in unphysical behaviour.

If we consider case~(\ref{case1}), evaporation into pure vapour, diffusion is excluded right from the beginning. As in case~(\ref{case2}), we assume that convective motion is very much faster than evaporation, what directly implies that the vapour density \(\rho \) is uniform in the entire gas layer. It is set by the conditions at the lateral boundaries of the gap. With other words, the vapour pressure is then a given constant and the governing thin-film equation in gradient dynamics form is Eq.~\eqref{eq:onefield:gov}. The corresponding energy is Eq.~\eqref{eq:vapourGap-energy} without the air energy and with constant \(\rho=\rho_\mathrm{vap}\), i.e.,
\begin{equation}
  F = \int_\Omega \left[ \frac{\gamma}{2} {(\nabla
      h)}^2 + g(h)
+h f_\mathrm{liq}(\rho_\mathrm{liq})
    +(d-h) f_\mathrm{vap}(\rho_\mathrm{vap})
    \right] \mathrm{d}^2 r,\label{eq:vapourGap-energy-case1} \end{equation}
Minimisation w.r.t.~film thickness gives
\begin{equation}
  \frac{\delta F}{\delta h} = -\gamma\Delta h+ g'(h) + f_\mathrm{liq} - f_\mathrm{vap},
  \label{eq:var1idcase1}
\end{equation}
i.e., the thin-film equation~\eqref{eq:onefield:gov} becomes
\begin{equation}
  \partial_t h \,=\,
  -\bnabla\bcdot\left\{Q\,\nabla [\gamma\Delta h - g'(h)] \right\}
  + M\left[\gamma\Delta h - g'(h) - f_\mathrm{liq} + f_\mathrm{vap} + p_\mathrm{vap}\right].
  \label{eq:onefield:gov-case1}
\end{equation}
Grouping all constants in the evaporation term into a single one, the equation corresponds to the standard thin-film description in the transition-limited case~\citep{Pism2004pre,Thie2010jpcm}.

\subsubsection{Specific functions and parameters}
\label{sec:tf-param}

One may now proceed by nondimensionalising, thereby using sensible
values of \(\rho_\mathrm{liq}\) and \(f_\mathrm{liq}\) as parameters. However, as they
are not independent, this may result in artefacts. The better approach
is to employ a specific equation of state / free energy that allows for liquid-gas phase transition and calculate these values for a particular temperature \(T\).

An example is the van der Waals equation of state, see e.g.,~\citet[§ 76 \& 84]{LandauLifshitz1987} giving the pressure

\begin{equation}
  p_\mathrm{vdW}=\frac{k_B T\rho}{1-b\rho}-a\rho^2
\end{equation}
where \(a\) and \(b\) are constants (attraction
strength and effective excluded volume, respectively). Approximated, in
the gas phase with \(a,b\to 0\) it becomes the ideal gas law \(p_\mathrm{id}=k_B T\rho \). The
corresponding Helmholtz free energy per volume is
\begin{equation}
  f_\mathrm{vdW}=-k_B T \rho\left[\log\left(\frac{1-\rho b}{\Lambda^3\rho}\right) + 1 \right]-a\rho^2
\end{equation}
becoming \(f_\mathrm{id}=k_B T \rho\left[\log\left(\Lambda^3\rho\right) - 1 \right]\)
for \(a,b\to 0\). Note that \(p=-f+\rho f'=\rho^2\partial_\rho(f(\rho)/\rho)\).

Further, the saturation vapour density \(\rho_\mathrm{sat}\) follows from the equilibrium condition, e.g., in the simple case of a thick flat liquid film (no Laplace and Derjaguin pressure) setting the evaporation rate to zero in Eq.~\eqref{eq:film-vapour-evolution-case2simple}:
\begin{equation}
  E = f_\mathrm{liq} - \rho_\mathrm{liq} k_B T \log\left(\frac{\rho_\mathrm{sat}}{\rho_\mathrm{tot}-\rho_\mathrm{sat}}\right) = 0.\label{eq:saturation-condition}
\end{equation}
The saturation vapour pressure \(p_\mathrm{sat}\) resulting from the equation of state may then be used to express the humidity as \(\phi = p / p_\mathrm{sat}\).

If one assumes a specific wetting potential \(g(h)\) that allows for partial wetting, a second spatially homogeneous equilibrium state is found typically at very small height \(h_a\) that corresponds to a microscopic adsorption layer
\begin{equation}
  g'(h_a) = \rho_\mathrm{liq}\,k_B T \log\left(\frac{\rho}{\rho_\mathrm{tot}-\rho}\right) - f_\mathrm{liq}.\label{eq:precursor-equilibrium}
\end{equation}
The height \(h_a\) depends on the vapour density \(\rho \) and will be slightly shifted from the height \(h_p\) where the Derjaguin pressure vanishes (\(g'(h_p) = 0\)), i.e., the adsorption layer height of the saturated case \(h_p = h_a(\rho_\mathrm{sat})\). In fact, this allows for a set of spatially homogeneous equilibrium states where the vapour density is an arbitrary constant \(0 < \rho \leq \rho_\mathrm{sat}\) controlled externally, e.g., by the humidity of the laboratory, while the substrate is macroscopically dry (the ``moist case'' of~\citet{DeG1985rmp}) with \(h = h_a\). Note that due to the saturation condition Eqs.~\eqref{eq:saturation-condition} and~\eqref{eq:precursor-equilibrium} we find \(g'(h_a) \leq 0\), i.e., the adsorption layer height is shifted to lower values \(h_a \leq h_p\).
When, on the other hand, an initial \(\rho(\boldsymbol{r}, t) \geq \rho_\mathrm{sat}\) is given, no equilibrium is guaranteed to exist (depending on the choice of the wetting potential). The evaporation rate \(E\) will then become negative leading to a transfer of material from the vapour phase to the liquid film, i.e., condensation.

In the following, we use the simple wetting potential for partially wetting liquids
\begin{equation}
  g(h) = \frac{H_A}{2h^2} \left(\frac{2 h_p^3}{5 h^3} - 1\right)\label{eq:wetting_potential}
\end{equation}
with the Hamaker constant \(H_A \approx \frac{5}{3}\gamma h_p^2 \theta_e^2\) that relates to the macroscopic equilibrium contact angle \(\theta_e\) by a mesoscopic Young's law: \(\cos \theta_e = 1 + \frac{1}{\gamma} g(h_p)\)~\citep{Chur1995acis}.

For the simulation of the thin-film model in section~\ref{sec:res-lw} we use the following set of realistic parameters (unless stated otherwise):
initial drop height $h_0 = h(r=0, t=0)=\SI{1}{mm}$,
gap height $d=\SI{3}{mm}$,
precursor film thickness $h_p$ = $h_0/1000$,
diffusion constant $D = \widetilde D k_B T = \SI{2.8e-5}{m^2/s}$~\citep{LeWi1954iec, MaMa1972jpcrd},
temperature $T=\SI{22}{\celsius}$,
contact angle $\theta_e = \SI{44.38}{\degree}$, and
a domain radius of $L=\SI{25}{mm}$.
The liquid properties are chosen according to water at room temperature: viscosity $\eta = \SI{8.9e-4}{\pascal \second}$, surface tension $\gamma = \SI{72e-3}{N/m}$, particle density according to mass density and molar mass $\rho_\mathrm{liq,mass} = \frac{\SI{997}{kg/m^3}}{\SI{18}{kg/mol}} N_A$, and the saturation vapour pressure $p_\mathrm{sat}=\SI{2643}{Pa}$. The parameter values are extracted from \citet{Lide2004}. The constant of phase change $M$ is employed as a free parameter that allows us to move between the different limiting cases.

\subsubsection{Radial symmetry and numerical implementation}
\label{sec:numerics}
For an efficient numerical simulation of the dynamics, we consider a radial symmetry, i.e., for the long-wave model \(h(\boldsymbol x, t) = h(r, t)\) and \(\psi(\boldsymbol x, t) = \psi(r, t)\). This allows us to reduce equations~\eqref{eq:film-vapour-evolution-case2simple} to a spatially one-dimensional model in polar coordinates
\begin{align}
  \partial_t h \, =       &
  \left(\partial_r + \frac{1}{r}\right) \left\{\frac{h^3}{3 \eta}\,\partial_r \left[-\gamma \left( \partial_r^2 h + \frac{1}{r} \partial_r h \right) + g'\left(h\right) \right]\right\}
  - j_\mathrm{evap}
  \nonumber                                                                                                         \\
  \partial_t[(d-h)\rho] = &
  \left(\partial_r + \frac{1}{r}\right) \left[D (d-h) \partial_r \rho\right]
  + \rho_\mathrm{liq} j_\mathrm{evap} \label{eq:film-vapour-evolution-case2simple-polar}
  \\
  \mathrm{where}\, \,
  j_\mathrm{evap}/M =     & -\gamma \left( \partial_r^2 h + \frac{1}{r} \partial_r h \right) + g'(h)+f_\mathrm{liq}
  - \rho_\mathrm{liq}k_B T \log\left(\frac{\rho}{\rho_\mathrm{tot}-\rho}\right).
  \nonumber
\end{align}
The dynamic equations are then solved using the finite-element method implemented in the C++ library \textit{oomph-lib}~\citep{HeHa2006}, which employs a second order BDF scheme for temporal integration. In particular, the library provides both spatial and temporal adaptivity, which is essential due to the multi-scale character of the dynamics, e.g., when spatially resolving the fields in the bulk and near the contact line. In the simulations, the domain is typically discretised with $\approx 1000$ grid points with spacings that range from $2^{-7}$ to $2^{-18}$ times the domain size.

In these calculations, we employ an adsorption layer of height \(h_p\) that is by a factor of \(10^3\) smaller than the initial drop height \(h(r=0, t=0)\). Larger ratios are possible but result in a strongly increasing numerical effort. As the effective adsorption layer height is coupled to the gas phase [see e.g.\ Eq.~\eqref{eq:precursor-equilibrium}], the overall pressure balance can cause liquid transport through the layer during equilibration. The resulting fluxes are negligible if the adsorption layer is very thin compared to the droplet size. The effect can be further suppressed by modulating the evaporation rate coefficient \(M(h)\) with a (smooth) step function, effectively disabling evaporation from the adsorption layer. This is in particular important on the slow timescale of droplet evaporation.

To ensure smoothness of all fields at the drop center ($r=0$), we employ natural (homogeneous) Neumann boundary conditions (BC)
\begin{align}
  \partial_r h\Big|_{r=0} = 0, \quad\quad \partial_r \frac{\delta F}{\delta \psi_1}\Big|_{r=0} = 0, \quad\text{and}\quad \partial_r [(d-h)\rho]\Big|_{r=0} = 0.\label{eq:thinfilm_bc_inner}
\end{align}
These conditions also ensure zero liquid and vapour flux through the computational domain boundary at $r=0$.
At the outer boundary far from the drop the gap between the plates is open, i.e., air and vapour can freely be exchanged with the surrounding laboratory. Thus, we assume a constant lab humidity (or vapour concentration) $\rho_\mathrm{lab}$ corresponding to a Dirichlet BC at $r=L$. Together with natural Neumann BC for the film profile $h$ and the chemical potential $\delta F / \delta \psi_1$, this gives the remaining BC
\begin{align}
  \partial_r h\Big|_{r=L} = 0, \quad\quad \partial_r \frac{\delta F}{\delta \psi_1}\Big|_{r=L} = 0, \quad\text{and}\quad \rho\Big|_{r=L} = \rho_\mathrm{lab}.\label{eq:thinfilm_bc_outer}
\end{align}
In particular, these conditions allow for a nonzero vapour flux through the outer boundary into the laboratory environment. The external vapour concentration is here chosen such that it corresponds to a low relative humidity of $\rho_\mathrm{lab}/\rho_\mathrm{sat} = \SI{10}{\percent}$.

Results obtained with the developed thin-film model are presented in Section~\ref{sec:res-lw}. In the subsequent Section~\ref{sec:res-stokes-exp} they are compared with Stokes-equation results and corresponding experiments.
\subsection{Stokes description}
\label{sec:stokes}
Next, we develop a description in terms of the Stokes equation for volatile liquids in the gap geometry again taking into account diffusion processes in the gas phase \textit{and} the mass-transfer at the liquid-gas interface. This will allow us to study the transition between transfer-limited and diffusion-limited cases also in the macroscale model. Then, important cases are compared to the thin-film model developed above.

The Stokes equations are solved on an axisymmetric domain, i.e., for the velocity field $\boldsymbol{u}(r,z,t)$ and the pressure $p(r,z,t)$, we solve
\begin{align}
  -\nabla p + \eta \bnabla\bcdot\left[\nabla\boldsymbol{u}+(\nabla\boldsymbol{u})^\text{t}\right] & =0 \\
  \bnabla\bcdot\boldsymbol{u}                                                                     & =0
\end{align}
in the liquid domain. At $r=0$ the conventional boundary conditions $u_r=0$ and $\partial_r p=0$ are imposed.
There is no precursor film considered in the Stokes description. Instead, a Navier slip boundary condition is used at the substrate at $z=0$. For comparison, the slip length is chosen to coincide with the precursor film thickness of the corresponding thin-film simulations, i.e.,
\begin{align}
  \frac{1}{h_p}u_r=\eta \left(\partial_z u_r + \partial_z u_z\right)\,,
\end{align}
see~\citet{SaKa2011el} for a comparison of precursor film and slip length models.

The liquid-gas interface is not described by a height function $h(r,t)$, but by a parametric representation $\boldsymbol{R}(s,t)$. On this interface, the surface tension is imposed as normal traction, i.e.,
\begin{align}
  \boldsymbol{n}\bcdot\left[-p\boldsymbol{1} + \eta \left(\nabla\boldsymbol{u}+(\nabla\boldsymbol{u})^\text{t}\right)\right] & =\left(-\gamma \kappa\right) \boldsymbol{n}\,,
\end{align}
where $\kappa=-\bnabla\bcdot \boldsymbol{n}$ is the full curvature and $\boldsymbol{n}$ is the normal vector.
The kinematic boundary condition has to consider the evaporation rate $j_\mathrm{evap}$, i.e. the relative normal motion between fluid velocity $\boldsymbol{u}$ and interface motion $\dot{\boldsymbol{R}}$ is enforced to be
\begin{align}
  \boldsymbol{n}\bcdot\left(\boldsymbol{u}-\dot{\boldsymbol{R}}\right) & =j_\mathrm{evap}\,.
\end{align}
The evaporation rate reads here
\begin{align}
  j_\mathrm{evap} / M & = p+f_\mathrm{liq}- \rho_\mathrm{liq} k_B T \log\left(\frac{\rho}{\rho_\mathrm{tot}-\rho}\right),
\end{align}
where the local pressure $p$ within the liquid appears instead of the terms $-\gamma \nabla^2 h+g'(h)$. In contrast to the thin-film model, where the Derjaguin pressure contribution $g'(h)$ can strongly influence the evaporation rate within the transition region between drop and precursor film, here, the singularity of the evaporation rate near the contact line is only limited by the finite mobility $M$ in the Stokes description with slip-length.

The contact angle $\theta$ is weakly imposed at the contact line by employing $\gamma(\cos\theta\:\boldsymbol{e}_r -\sin\theta\:\boldsymbol{e}_z)$ as surface tension force at the contact line, which balances when the free surface attains the prescribed contact angle $\theta$.

The vapour diffusion in the gas phase is also fully resolved in the direction orthogonal to the substrate. Thus, instead of solving for the evolution of the height-averaged vapour density $\rho(r,t)$, the diffusion equation
\begin{align}
  \partial_t \rho=D\nabla^2 \rho
\end{align}
is solved for the density $\rho(r,z,t)$.

No-flux conditions are used at the top plate, at the axis of symmetry and at the substrate-gas interface, which ranges from the contact line at $(r_\text{c},0)$ to the end of the domain at $(L,0)$, and lab humidity is imposed at the far end, i.e.
\begin{align}
  \partial_r \rho\Big|_{r=0} =0\,,\qquad  \partial_z \rho\Big|_{z=d} =0\,,\qquad  \partial_z \rho\Big|_{r>r_c,z=0} =0\,, \qquad \rho\Big|_{r=L}=\rho_\mathrm{lab}\,.
\end{align}
Finally, at the liquid-gas interface, the evaporation flux is considered:
\begin{align}
  -D\nabla \rho \bcdot \boldsymbol{n}=\rho_\mathrm{liq}\,j_\mathrm{evap}\,.
\end{align}

The equations are implemented in \textit{oomph-lib} employing a sharp-interface Arbitrary Lagrangian-Eulerian method, i.e. the mesh nodes are moving in such  a way that the liquid-gas interface is always conforming with the mesh. Mesh reconstruction and subsequent interpolation of all fields to the new meshes is invoked whenever the mesh distortion exceeds given thresholds. The mesh has enhanced resolution at the contact line and the typical number of degrees of freedom of the discretised system is chosen to be around 1,000,000.

\subsection{Experiments}
\label{sec:exp}

\begin{figure}
  \centering
  \includegraphics[width=1\hsize]{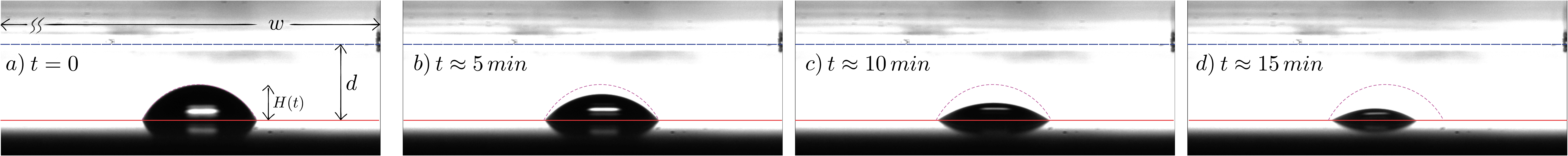}
  \caption{An example of experiments of sessile droplet evaporation in a gap. The red solid line is the bottom plate. The blue dashed lines shows the top plate. The initial shape of the droplet is shown with a dotted line in all panels. For this test, the initial diameter of the drop was \SI{1.33}{mm} and the gap size was $d=\SI{0.88}{mm}$.}\label{fig:exper}
\end{figure}

Experiments are done using a simple shadowgraphy setup. The millimetric water droplet (volume of $\approx \SI{0.4}{\micro l}$) is placed on a cover glass with a thickness of $\approx \SI{0.2}{mm}$ (initial contact angle of $\approx \SI{65}{\degree}$) - see Fig.~\ref{fig:exper}. The cover glass is placed on top of a holder such that the air occupies the space below it. The top surface is a thick plastic circular disk with a diameter ($w$) much larger than the size of the droplet. The gap between the plates ($d$) is accurately controlled with a micrometer positioning stage, and special care is taken to make sure both top and bottom surfaces are horizontal. An LED light source illuminates the test section, and the images of the droplet are recorded using a CMOS camera attached to a zoom lens (see Fig.~\ref{fig:exper} for example snapshots). Experiments are performed for different levels of confinement, varying the gap height $d$. The changing shape of the droplet over time is acquired by the standard processing of the images.

\section{Results of thin-film model}
\label{sec:res-lw}
\begin{figure}
  \centering
  \includegraphics[width=\textwidth]{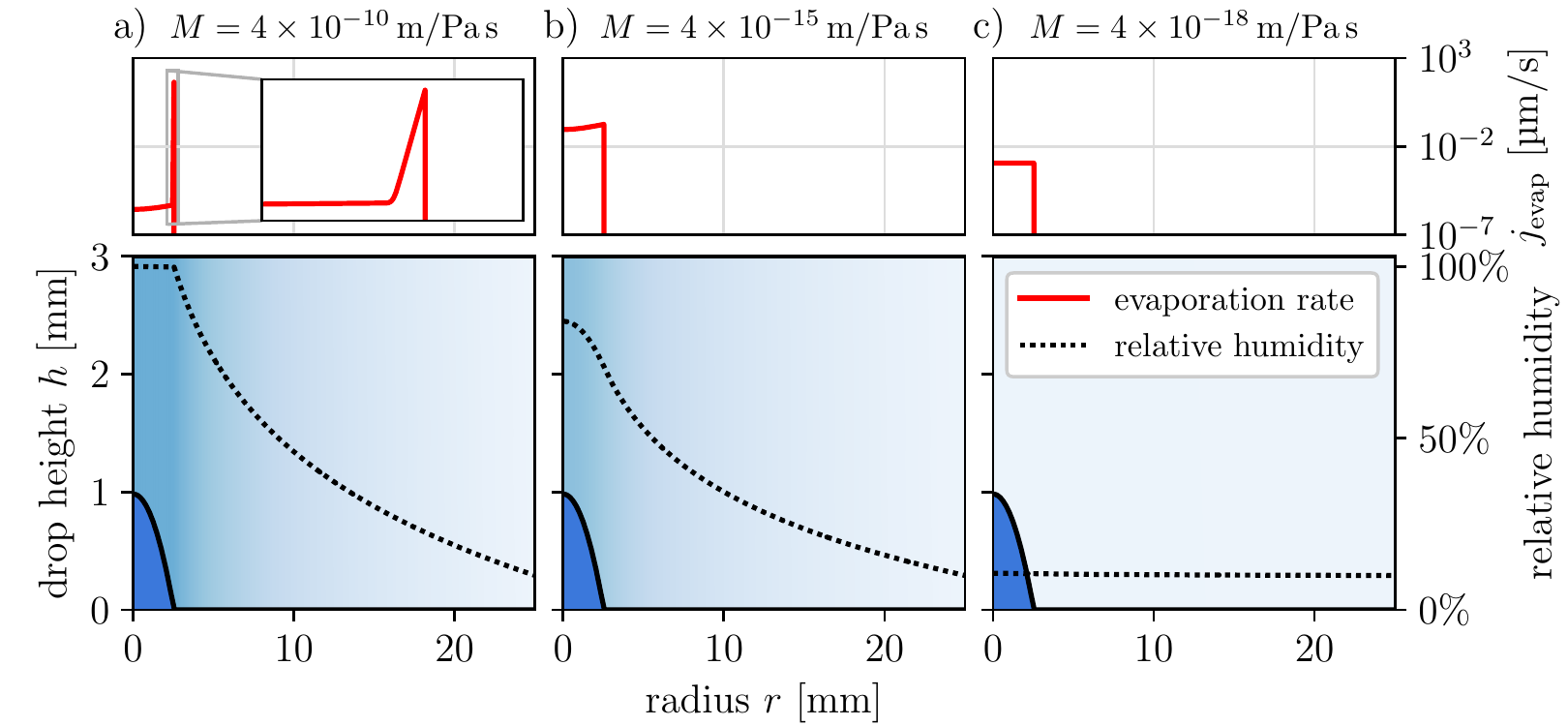}
  \caption{Single simulation snapshots are shown obtained with the thin-film model~\eqref{eq:film-vapour-evolution-case2simple-polar}. Panel~(a) gives the diffusion-limited case, (b) an intermediate case, and (c) the phase transition-limited case at evaporation rate constant $M$ as given above the panels. The gap height is fixed at $d=\SI{3}{mm}$. All other parameters are identical and correspond to the values given at the end of Section~\ref{sec:tf-param}. The top panels show the local evaporation rate $j_\mathrm{evap}(r,t)$ while the bottom panels give the droplet profile (solid line, filled blue, left hand side scale) and the local relative humidity (dotted line and bluish shading, right hand side scale). The inset in the top panel (a) magnifies the peak in evaporation rate at the contact line (lateral size \SI{0.7}{mm}).}\label{fig:thinfilm_snapshots}
\end{figure}

First, we employ the thin-film model in radially symmetric form~\eqref{eq:film-vapour-evolution-case2simple-polar} with BC~\eqref{eq:thinfilm_bc_inner} and~\eqref{eq:thinfilm_bc_outer} to simulate an evaporating droplet for different modi of evaporation. Figure~\ref{fig:thinfilm_snapshots} gives an overview by comparing single snapshots for the cases of diffusion-limited [Fig.~\ref{fig:thinfilm_snapshots}~(a)] and phase transition-limited [Fig.~\ref{fig:thinfilm_snapshots}~(c)] mass transfer from the drop to the gas atmosphere. An intermediate case is also given [Fig.~\ref{fig:thinfilm_snapshots}~(b)]. The three cases are obtained by only changing the value of the evaporation rate constant $M$, while all other parameters are fixed. All simulations were initialised with a sessile droplet profile of \SI{1}{mm} height, representing a droplet equilibrated without evaporation, in a homogeneous atmosphere of a constant (low) humidity. The snapshots are taken after initial transients have passed and a (quasi-static) vapour concentration profile in the gas phase has been established that subsequently changes on a much slower time scale.

In particular, the bottom panels show the drop profile (solid black line, right hand side scale), the vapour concentration (i.e., relative humidity) as bluish background shading and also as black dotted line (left hand side scale). The upper panels give the corresponding evaporation rate profiles on a logarithmic scale. The left column [Fig.~\ref{fig:thinfilm_snapshots}~(a)] presents the diffusion-limited case at relatively large $M=\SI{4e-10}{m/Pa\,s}$, i.e., the diffusive transport of vapour through the gas phase is much slower than the phase change, thereby effectively controlling the entire process. Note that above the droplet the vapour is nearly saturated due to the gap-geometry. Starting in the contact line region the concentration logarithmically decays towards the edge of the plates where the humidity is kept at 10\%. The analytical form is discussed below in appendix~\ref{sec:res-limit}.

Remarkable is the clearly visible divergence of the evaporation rate when approaching the contact line from inside the drop [top panel of Fig.~\ref{fig:thinfilm_snapshots}~(a)]. In this region the local evaporation rate changes exponentially by about seven orders of magnitude leading to a very sharp spike. See Fig.~\ref{fig:thinfilm_snapshot_zoom} in the appendix for a highly zoomed version of the peak demonstrating the smoothness of the solution in the vicinity of the contact line.

In contrast, Fig.~\ref{fig:thinfilm_snapshots}~(c), i.e., the right column, shows the transition-limited case at relatively small $M=\SI{4e-18}{m/Pa\,s}$, i.e., the diffusive transport in the gas phase is much faster than the phase change. Now it is the latter that controls the entire process. As diffusion is fast, all surplus vapour is rapidly transported out of the gap between the plates and the vapour concentration profile remains nearly homogeneous at lab humidity. In consequence, the evaporation rate is nearly constant along the entire surface of the drop and rapidly falls to zero in the contact line region [top panel of Fig.~\ref{fig:thinfilm_snapshots}~(c)].

Beside the two limiting cases respectively considered by the two groups of thin-film models in the literature, our model is also able to simulate 'mixed cases' where the time scales of the involved processes are not strongly separated. An example is presented in the centre column of Fig.~\ref{fig:thinfilm_snapshots} at moderate $M=\SI{4e-15}{m/Pa\,s}$. Here, diffusion is fast enough to keep the concentration at the drop centre below saturation. It is also sufficiently slow for a decaying concentration profile to develop outside the drop. Accordingly, also the profile of the evaporation rate [top panel of Fig.~\ref{fig:thinfilm_snapshots}~(b)] shows features of both limiting cases: the flux is moderately large above the drop, increases by about half an order of magnitude towards the contact line region where it steeply decays to zero.

\begin{figure}
  \centering
  \includegraphics[width=\textwidth]{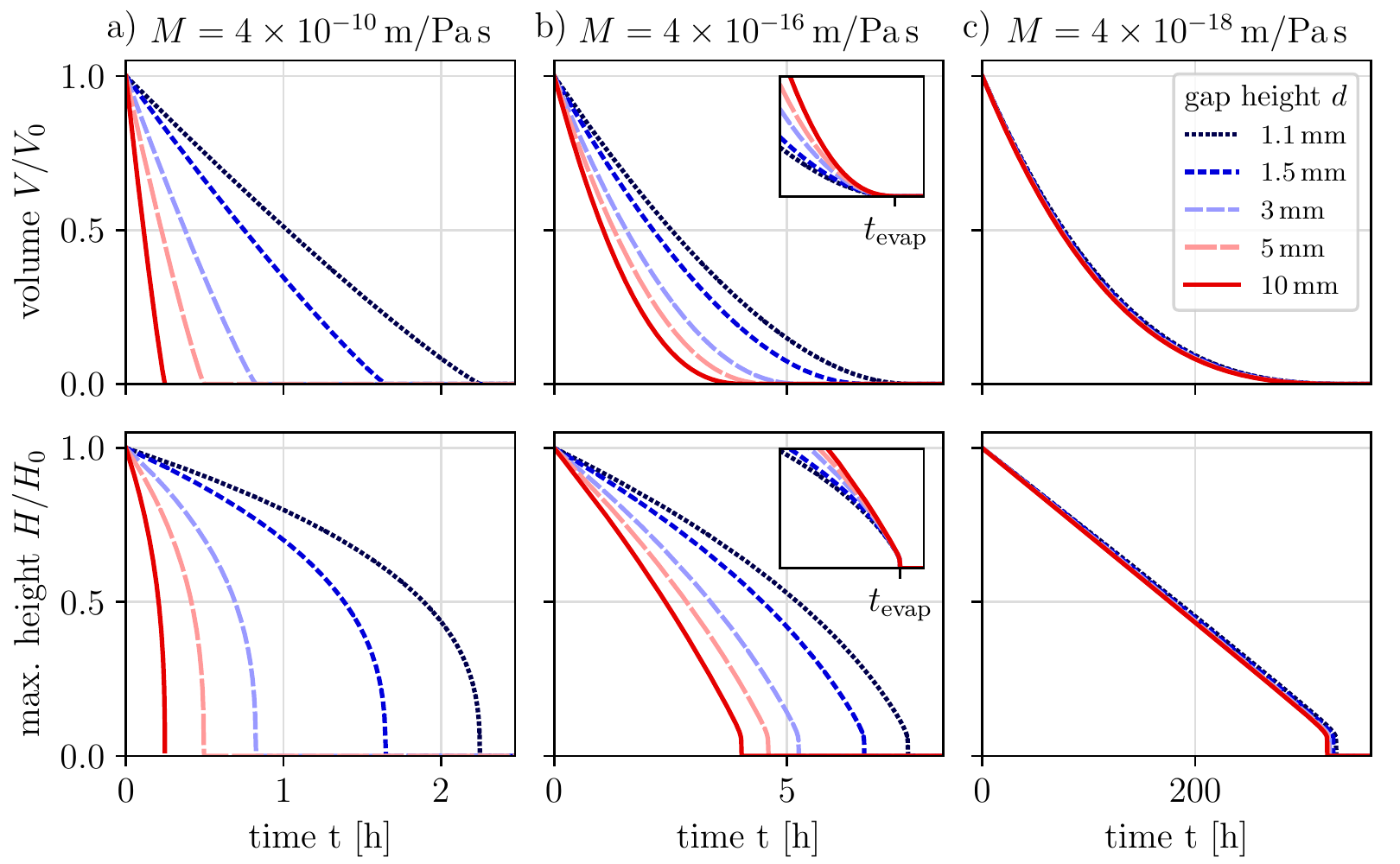}
  \caption{The thin-film model is used to determine the change in (top) drop volume and (bottom) drop height for various gap heights $d$ as given in the legend in (a) the diffusion-limited case, (b) an intermediate case, and (c) the phase transition-limited case via different values of the phase transition coefficient $M$. The inset in (b) shows the same data but shifted in time such that the curves coincide at the time the drop has fully evaporated, revealing that the curves show the same phase transition-limited behaviour when the droplet is much smaller than the gap height $d$.}\label{fig:thinfilm_volume_height}
\end{figure}

After having established the stark difference in evaporation flux and concentration profiles seen for single snapshots in the two limiting cases, next we consider how the time evolution differs between them. To do so, in Fig.~\ref{fig:thinfilm_volume_height} we study how drop volume and height change in time for a number of different gap widths $d$ in the diffusion-limited and the phase transition-limited case. In the former case [Fig.~\ref{fig:thinfilm_volume_height}~(a)], the drop volume shows a linear decrease in time. Thereby the rate is proportional to the gap width $d$ as it directly controls the overall vapour flux in our height-averaged setting (cf.~Appendix~\ref{sec:res-limit}).
Correspondingly, the drop height shrinks with a power law $\sim (t_\mathrm{evap}-t)^{1/3}$ as expected for a nearly constant contact angle. Here, $t_\mathrm{evap}$ is the time when the complete drop has evaporated.

In contrast, in the transition-limited case [Fig.~\ref{fig:thinfilm_volume_height}~(c)], the drop height decreases linearly with a rate nearly independent of the gap width. Correspondingly, the drop volume shrinks with a power law $(t_\mathrm{evap}-t)^{3}$. Note that in the final stage when the drop is rather small, the remaining volume is rapidly absorbed into the adsorption layer. This is a mesoscale effect resulting from the dominance of the wetting energy for very small drop size. We do not further discuss it here.

Interestingly, in the intermediate case of moderate evaporation [Fig.~\ref{fig:thinfilm_volume_height}~(b)], features of both limiting cases are visible: in contrast to the transition-limited case, the evolution depends on gap width, but less so than seen in the diffusion-limited case. Neither the drop volume nor the drop height decrease linearly. Instead, in the course of the time evolution they cross over from a diffusion-limited start, where slopes of approximately linear decays in volume are proportional to gap width, to a transition-limited end with cubic decrease of volume and final drop collapse.

In the transition-limited case, the phase transfer rate decreases with the drop size, as it is limited by its surface area. Note that this is not true for nanoscopic droplets, where the Laplace pressure  significantly contributes to the drop chemical potential, causing an increased evaporation, i.e., the Kelvin effect. This is not observed in our mesoscale simulations, due to the size of the droplets. Nevertheless, we expect that the behaviour of very small droplets should always tend towards the transition-limited behaviour because their surface area shrinks while the rate at which lateral diffusion transports vapour particles away remains (nearly) constant. The magnification in the insets of Fig.~\ref{fig:thinfilm_volume_height}~(b) therefore depicts the measured volume and height data shifted in time such that the curves coincide at the time $t_\mathrm{evap}$ of complete evaporation. There, in particular at large gap heights $d$, i.e., when the diffusion is less dominant, for small drop sizes the behaviour converges to a common curve that resembles the transition-limited case of Fig.~\ref{fig:thinfilm_volume_height}~(c).

Note that in all considered cases evaporation is sufficiently slow to only see a minor evaporation-induced difference \citep{Morr2001jfm,ToTP2012jem,ReCo2017prf} below 2\% between the observed quasi-steady contact angle and the equilibrium angle. Further, the contact line smoothly recedes as no contact line pinning occurs for the assumed ideally homogeneous and smooth substrate, i.e., there is no contact angle hysteresis. This aspect will be refined in the following two subsections.

\subsection{Dependence on the contact angle}
Before we investigate the effects of contact angle hysteresis, we check the overall influence of the contact angle on the dynamics. We therefore perform long-wave simulations of droplets with varying equilibrium contact angle, i.e., we control the parameter $\theta_e$ in the wetting potential (Eq.~\eqref{eq:wetting_potential}) and initialise the simulation with a droplet of according equilibrium shape. Droplets of identical volume but different radii should then show different evaporation rates merely due to their differences in surface area, which masks the actual effect of the contact angle. Hence, in the following we consider droplets of identical initial radius and accept that they have various initial volumes, depending on their contact angle. Naturally, a drop of smaller volume evaporates faster than a large drop, again hiding the mere influence of the contact angle. Yet, we can account for the different drop volumes by normalizing the time scale with the initial volume $\tilde t = t / V(t=0)$.

\begin{figure}
  \centering
  \includegraphics[width=\textwidth]{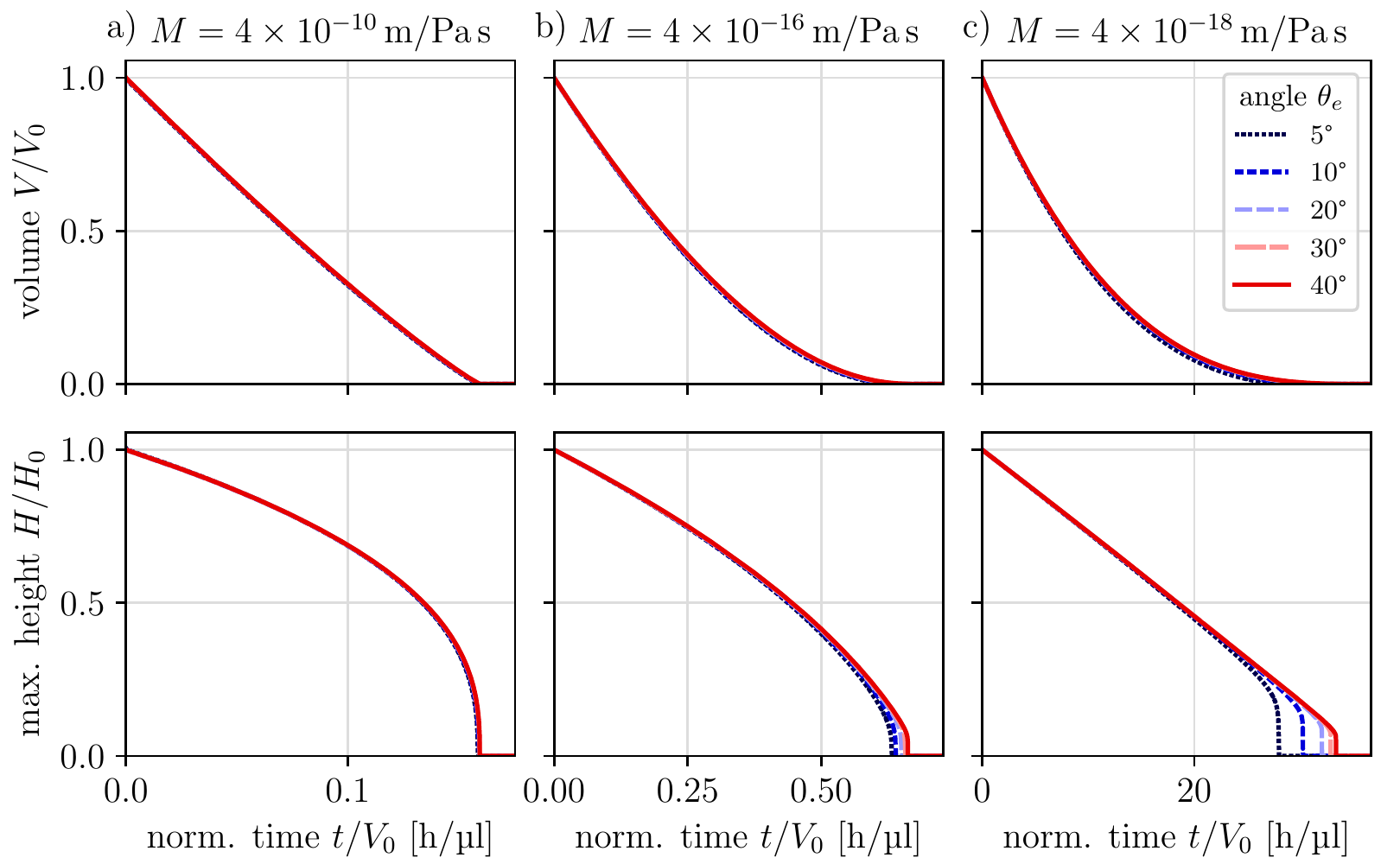}
  \caption{Droplet volume and height over time for long-wave simulations identical to Fig.~\ref{fig:thinfilm_volume_height} but with varying values of the contact angle parameter $\theta_e$. The gap height is now fixed at $d=\SI{1.5}{mm}$. All drops were initialised with their respective contact angle, same radius but varying initial volume $V_0$. The normalised time scale $t/V_0$ accounts for the different evaporation times of the different drop volumes, revealing that the contact angle itself has little to no qualitative effect on drops of the same radius.}\label{fig:thinfilm_contact_angle_variation}
\end{figure}

We then employ measures identical to the ones in Fig.~\ref{fig:thinfilm_volume_height}, but vary the contact angle $\theta_e$ instead of the gap height, which now remains fixed at $d=\SI{1.5}{mm}$. All other parameters are kept at the same values as given at the end of Section~\ref{sec:tf-param}. In Fig.~\ref{fig:thinfilm_contact_angle_variation} we plot the measured drop volume $V(t)$ and  height $H(t)$ versus the normalised time $t = t / V(t=0)$ for three different values of the phase transition coefficient $M$, again accounting for the diffusion-limited case (a), the phase transition-limited case (c), and an intermediate case (b).

Notably, in all configurations the dynamic behaviour is mostly independent of the contact angle. The only visible differences occur when the droplets become very small. This is most pronounced for low contact angles and in the final drop collapse stage of the transition-limited case. This is again a mesoscale effect due to the dominant wetting potential for droplets of low film heights. The otherwise coinciding curves in Fig.~\ref{fig:thinfilm_contact_angle_variation} demonstrate that all effects of the contact angle on the evaporation rate are due to changes in the liquid surface area. In particular, low contact angle droplets have a higher surface area than a droplet with higher contact angle of the same volume and will evaporate faster, depending on the relation of the total evaporation rate to the surface area. See the appendix~\ref{sec:res-limit} for an asymptotic approach to determining the latter. There, we find that the total evaporation rate scales as $1/R^2$ with the drop radius in the phase transition-limited case and is mostly independent of it in the diffusion-limited case, i.e., the behaviour is also independent of the contact angle.

\subsection{Effects of contact line pinning}\label{sec:pinning}
In all simulations discussed up to here, the contact line continuously recedes during evaporation in order to retain the contact angle governed by the wetting potential $g(h)$ [Eq.~\eqref{eq:wetting_potential}]. However, our choice of a simple wetting potential does not account for contact angle hysteresis, e.g., due to roughness of the solid substrate. Hence, the previous discussion does not include any effects of pinned contact lines.

Inhomogeneities of the wettability are, however, easily incorporated into the formalism through a spatial modulation of the wetting potential~\citep{TBBB2003epje}. In the following, we simulate a contact angle hysteresis by modulating $g(h)$ with a (possibly smoothed) step function $\xi(r)$ that drastically increases the wettability (by lowering the energy at the minimum of the wetting potential) in the substrate region of the initial droplet radius $R_0$. Namely,
\begin{equation}
  g(h, r) = \xi(r)\,g(h) \quad \text{with} \quad \xi(r) = \begin{cases}
    1   & \text{for}~~r\geq R_0 \\
    0.1 & \text{for}~~r< R_0.
  \end{cases}
\end{equation}
This corresponds to a contact angle hysteresis with the threshold angles $\theta_{\min}=\sqrt{0.1}\,\theta_{\max}$, i.e., between \SI{14.03}{\degree} and \SI{44.38}{\degree} in our simulations.

\begin{figure}
  \centering
  \includegraphics[width=\textwidth]{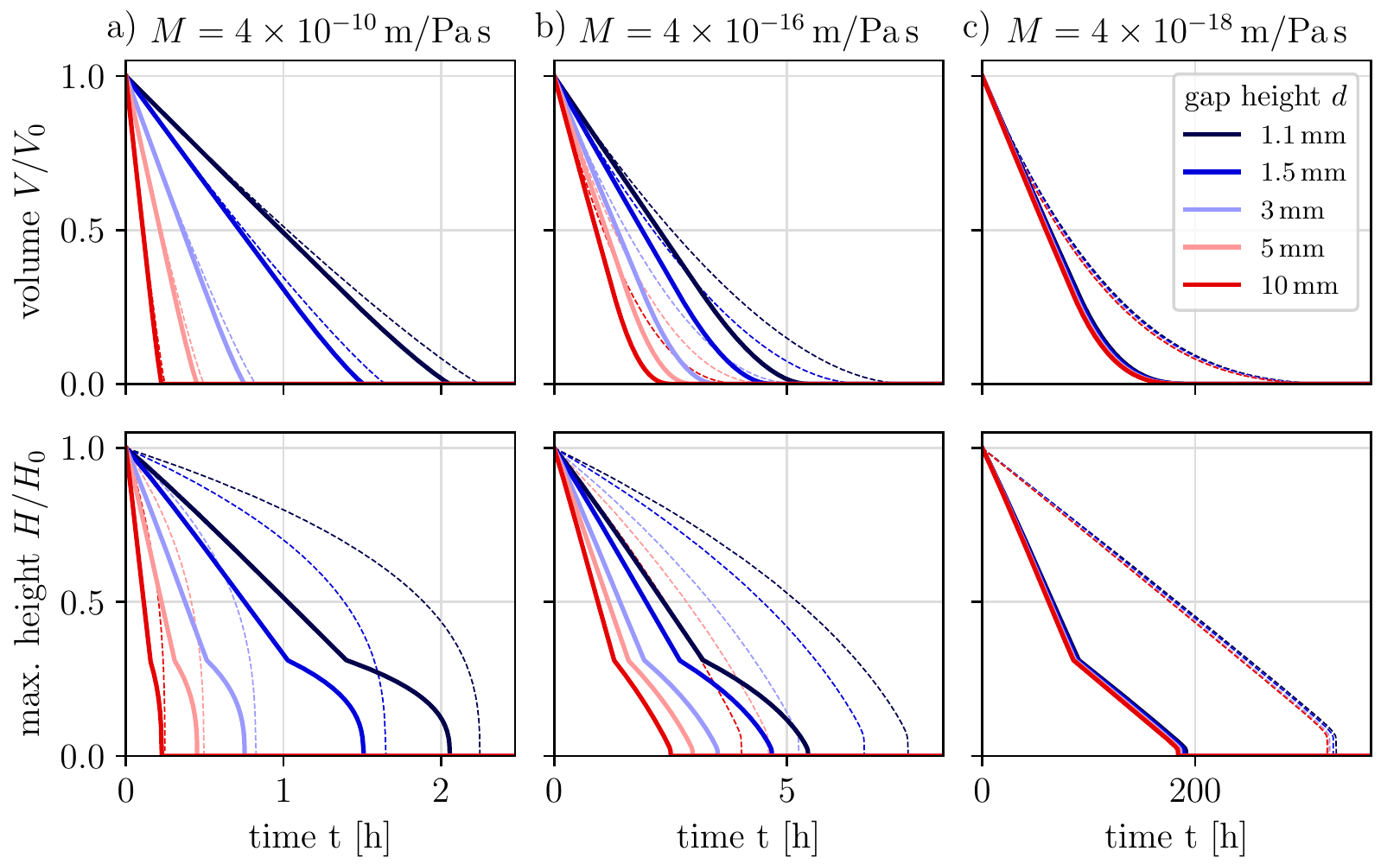}
  \caption{Droplet volume and height over time for long-wave simulations identical to Fig.~\ref{fig:thinfilm_volume_height} but with an initial pinning of the drop's contact line due to a contact angle hysteresis of $\SI{30}{\degree}$. For comparison, the simulations without pinning are included as thin dashed lines. The noticeable corner in the height curves indicate the depinning event. While the contact line is pinned, the volume evaporation rate $\mathrm{d}V/\mathrm{d}t$ remains nearly constant, accelerating evaporation particularly in the transition-limited case.}\label{fig:thinfilm_pinning}
\end{figure}

In Fig.~\ref{fig:thinfilm_pinning} we provide a comparison between the long-wave simulations without pinning of the contact line, namely, the results of Fig.~\ref{fig:thinfilm_volume_height}, with simulations including contact angle hysteresis. Apparently, independently of the considered parameter regime, the drop volume initially decreases linearly, i.e., the evaporation rate appears constant. Note that here as well the drop height seems to decrease nearly linearly, because the volume $V$ and height $H$ of a shallow drop of fixed base radius $R$ are related as $V=\pi/2\,HR^2$. At a specific height, when the droplet has reached the lower threshold contact angle $\theta_{\min}$ the contact line depins. This causes a distinct corner in the $H(t)$ curve as the linear relation between height and volume gives place to the usual cubic relation $V \sim R^3$ in the diffusion-limited case (Fig.~\ref{fig:thinfilm_pinning}~(a)) and a linear relation with a different slope in the transition-limited case (Fig.~\ref{fig:thinfilm_pinning}~(c)).

The constant evaporation rate in the case of a pinned contact line also appears in the approximate expressions of the total evaporation rate derived in appendix~\ref{sec:res-limit}. There, the evaporation rates for both, the strongly diffusion-limited and strongly transition-limited case, depend only on the radius $R$, but not on any other characteristics of the drop shape. In the transition-limited case, the evaporation rate is quite uniform over the drop surface ($\sim R^2$ in the long-wave limit), whereas in the diffusion limited case, evaporation mostly occurs in the vicinity of the contact line, i.e., it scales with the droplet radius $R$.

Once the contact line has depinned, the droplet continues to shrink with a contact angle close to $\theta_{\min}$, where the curves for volume $V(t)$ and height $H(t)$ recover the characteristic shape of the case without contact angle hysteresis studied above. Note, in particular, that their slope matches exactly for the transition-limited case in Fig.~\ref{fig:thinfilm_pinning}~(c).

Overall, due to the constant rate of evaporation, in all cases pinning results in a faster drop evaporation. The effect is, however, most pronounced in the phase transition-limited case, where the evaporation rate exhibits a dependency on the drop surface area $\sim R^2$. The effects are more subtle for the strongly diffusion-limited case, where the drop shape is mostly irrelevant as the vapour diffusion is the limiting factor.

\section{Comparison to Stokes description and experiments}
\label{sec:res-stokes-exp}

\begin{figure}
  \centering
  \includegraphics[width=0.999\textwidth]{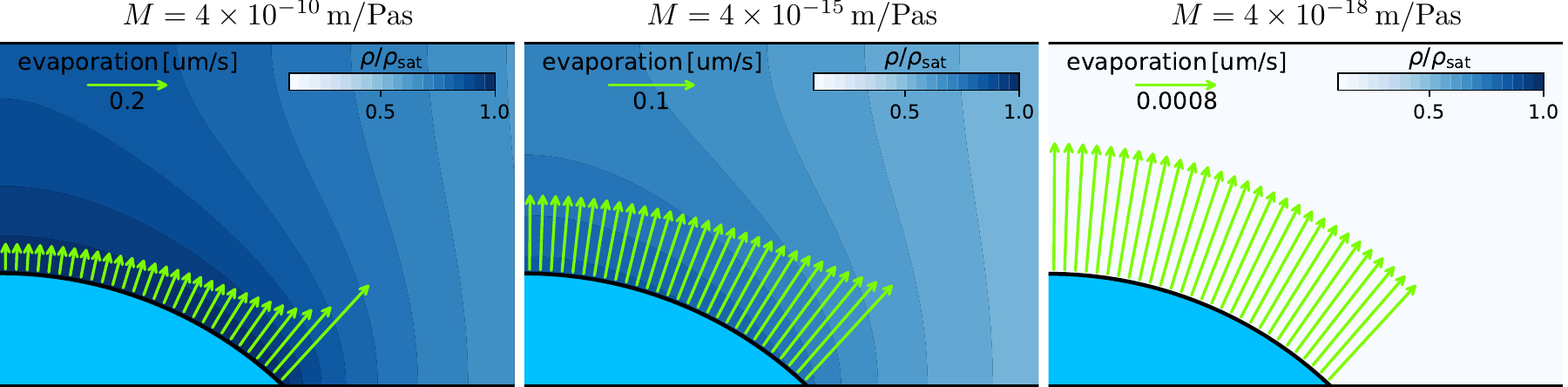}
  \caption{Simulation snapshots obtained using the Stokes model at identical parameters and times as employed in Fig.~\ref{fig:thinfilm_snapshots} for the thin-film model. Panel~(a) gives the diffusion-limited case, (b) an intermediate case, and (c) the phase transition-limited case. The evaporation rate is indicated by the arrows at the liquid-vapour interface, whereas the vapour density $\rho(r,z,t)$ is given by the background shading.}\label{fig:stokes_snapshots}
\end{figure}

Next, we compare the presented results obtained with the developed thin-film model to simulations with the Stokes model in a gap geometry and to experiments with evaporating sessile drops in a narrow gap.
In Fig.~\ref{fig:stokes_snapshots}, snapshots obtained with the Stokes model are shown, at parameters and times that exactly coincide with the ones employed in Fig.~\ref{fig:thinfilm_snapshots} for the thin-film model. In the diffusion-limited case, Fig.~\ref{fig:stokes_snapshots}~(a), the evaporation rate is highest near the contact line. However, in contrast to the thin-film model, where the Derjaguin pressure $g'(h)$ has an influence on the evaporation rate, this corresponding effect is reflected in the Stokes model by the change from drop to bare substrate. With other words the jump from a finite evaporation rate at the drop edge to zero at the bare substrate is slightly smoothed by the Derjaguin pressure in the thin-film model. A stronger difference can be seen in the vapour distribution near the droplet apex that here clearly shows a deviation from the $z$-independent distribution assumed in the thin-film model. However, a distance away from the droplet, the vapour distribution indeed becomes more and more uniform in the direction vertical to the substrate, i.e., diffusion turns into a simple one-dimensional radial process as also encoded in the thin-film model.

The intermediate case depicted in Fig.~\ref{fig:stokes_snapshots}~(b) shows similar features as the corresponding thin-film model result in Fig.~\ref{fig:thinfilm_snapshots}~(b), i.e. an almost uniform evaporation rate with a slight enhancement near the contact line, where the vapour diffusion still contributes as limiting factor, i.e. vapour gradients are still visible.

The rate-limited case in Fig.~\ref{fig:stokes_snapshots}~(c) features a uniform evaporation rate and a vapour diffusion that is sufficiently fast to approach an almost homogeneous vapour concentration fixed by the lab vapour concentration. This is in full agreement with the corresponding results of the thin-film model in Fig.~\ref{fig:thinfilm_snapshots}~(c).

\begin{figure}
  \centering
  \includegraphics[width=\textwidth]{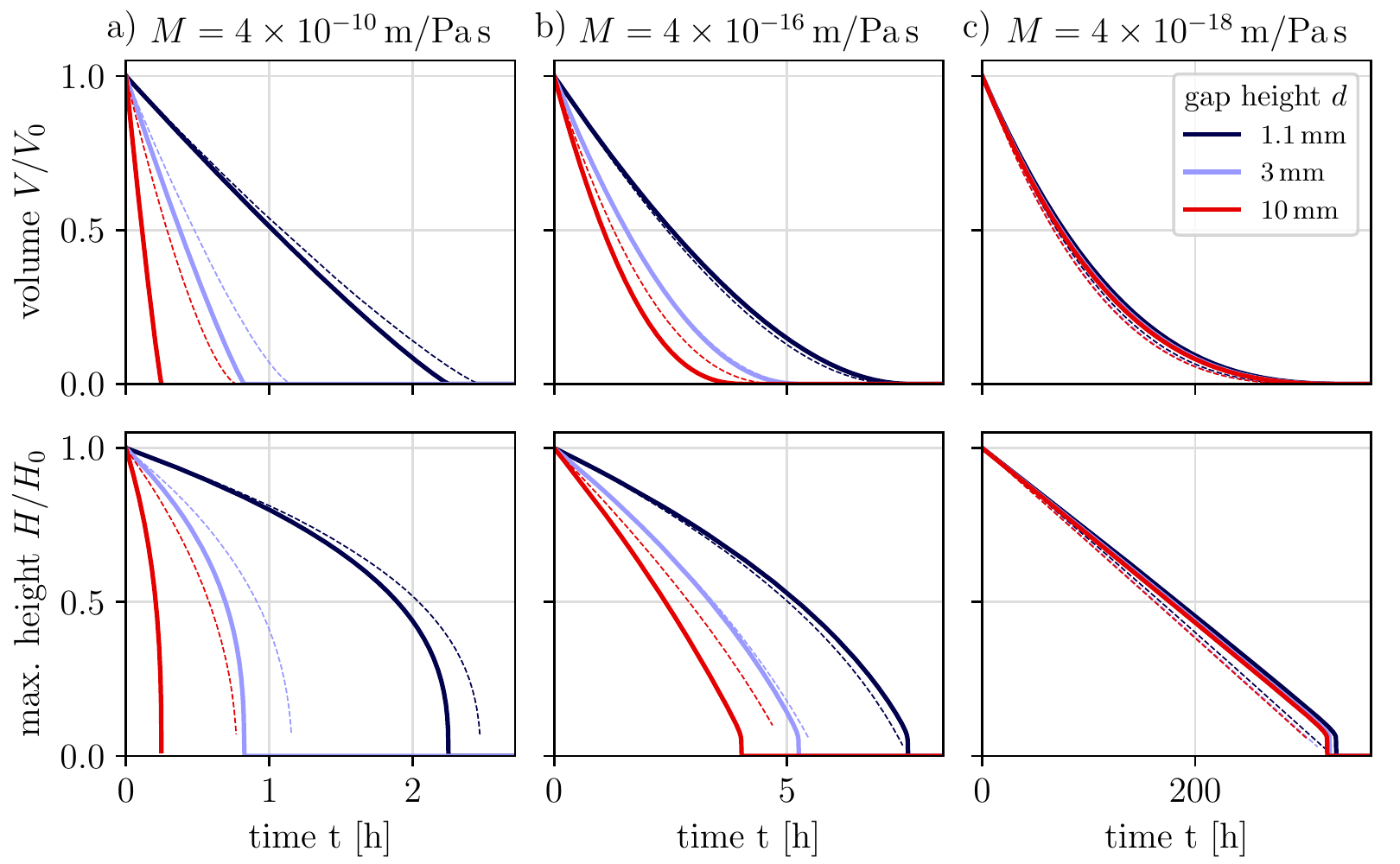}
  \caption{Comparison of the volume evolution $V(t)$ and the height evolution $h(t)$ for the thin-film model (solid lines) and the Stokes model (dashed lines) for (a) the diffusion-limited case, (b) an intermediate case, and (c) the phase transition-limited case.}\label{fig:stokes_volume_height}
\end{figure}

Next we assess the influence of the approximations made in the mesoscopic model in the shallow-drop case, i.e., assuming a $z$-averaged vapour distribution, a simplified expression for the curvature resulting in parabolic drop profiles, and laminar flow. To do so Fig.~\ref{fig:stokes_volume_height} gives the volume and height evolution for both models. Note that, due to the parabolic drop profile in the thin-film model and the spherical-cap shaped profile in the Stokes model, the equilibrium contact angle of \SI{44.38}{\degree} (as used for the thin-film results) had to be slightly reduced to \SI{43.38}{\degree} in the Stokes description to simultaneously have identical initial heights and volumes in the two models.

Obviously, for the diffusion-limited case in Fig.~\ref{fig:stokes_volume_height}~(a), the agreement is better for smaller gap heights, i.e., for smaller distances between the droplet and the upper plate. For $d=\SI{1.1}{\milli\meter}$, results in the initial stage of the evolution almost coincide, whereas they deviate towards the final stage, when the droplet height $h$ becomes considerably smaller than the plate distance $d$. The diffusive vapour transport in the Stokes model is slower than predicted by the purely radial transport assumed in the thin-film model. The larger the initial distance of drop apex to upper plate, the higher the initial deviation. For the largest considered plate distance of $\SI{10}{\milli\meter}$, the assumption of a $z$-averaged vapour distribution underpredicts the total evaporation time by a factor of $3$.

In the intermediate case, Fig.~\ref{fig:stokes_volume_height}~(b), it is apparent that at small gap widths the Stokes model shows nearly identical but slightly faster evaporation than the thin-film model. Since such slightly faster evaporation is also visible for the transition-limited case in Fig.~\ref{fig:stokes_volume_height}~(c), it can be attributed to the different mathematical treatment of the free surface: in the shallow-drop case, the evaporation rate is effectively determined by the base area $\pi R^2$ of the parabolic drop.
In the Stokes model, the full surface area of a spherical drop is considered, leading to an increased total volume loss rate once the transfer rate becomes the limiting factor.
Note that this discrepancy disappears in the full-curvature formulation of the mesoscopic model, implying improved expressions for Laplace pressure and evaporation terms. For large plate distances $d$ in the intermediate case, however, our central approximation of a $z$-averaged vapour density again results in an underprediction of the total evaporation time in the mesoscopic model.

The above argument is further confirmed by the transition-limited case in Fig.~\ref{fig:stokes_volume_height}~(c). There, the macroscopic Stokes calculations uniformly give a slightly faster evaporation than in the mesoscopic model for all considered gap heights. This indeed implies that the cause of the difference is in the treatment of the drop surface area as discussed before. In the full-curvature formulation, mesoscopic and macroscopic approaches fully coincide.

We conclude that as expected the agreement of the results obtained with the developed mesoscopic approach and the macroscopic Stokes approach improves for smaller gap heights $d$, as then the gas phase becomes increasingly vertically homogeneous also in the Stokes simulations. Similarly, smaller contact angles $\theta_e$ allow for lower drop heights and thus for lower gap width, therefore improving the agreement.

Finally, we compare the results obtained from the Stokes and thin-film simulations to the experiments.
Therefore we adapt the parameters and initial condition of the simulations such that they closely match the experiment. The resulting parameters are the same as mentioned in section~\ref{sec:tf-param}, with the only difference that we here assume a relative humidity of $\rho_\mathrm{lab}/\rho_\mathrm{sat}=\SI{50}{\percent}$ at the far end of the gap (at $r=\SI{25}{mm}$).
Additionally, in the simulations the phase transition coefficient is chosen as $M=\SI{4e-10}{m/Pa\,s}$, i.e., the process is limited by diffusion rather than phase change.
The experimental drop shape and volume were extracted from the shadowgraphy images and used to initialise the simulations with droplets of identical size (see table~\ref{tab:exp_drop_parameters} in the appendix for the measured initial drop shape parameters).
Since in the experiment the droplet exhibits a strong pinning effect at its initial contact line position and depins only at a very low contact angle (see also Fig.~\ref{fig:exper}), we here pin the contact line in the thin-film model using the mechanism described in section~\ref{sec:pinning} and replace the Navier slip condition with a no-slip condition in the Stokes model.

\begin{figure}
  \centering
  \includegraphics[width=0.95\textwidth]{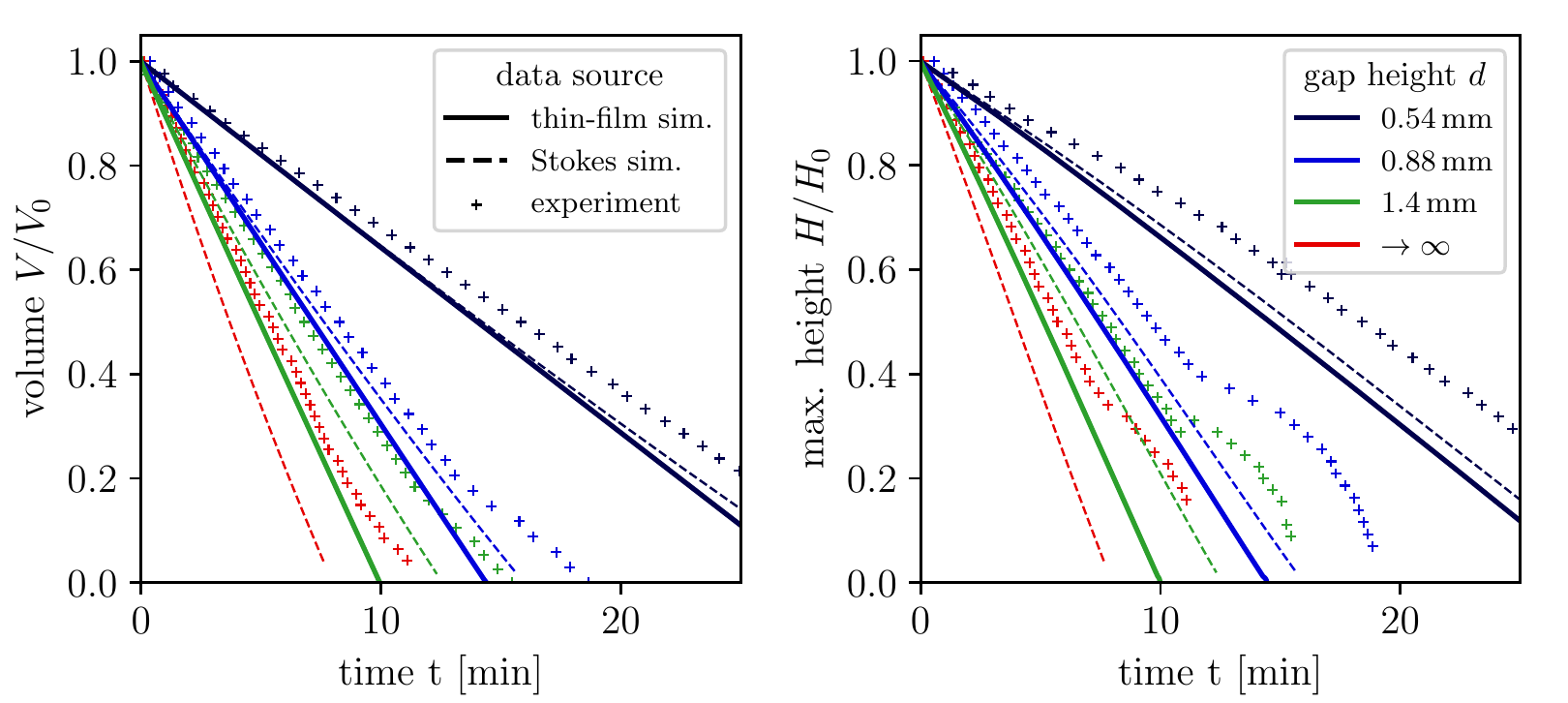}
  \caption{Comparison between experiments (crosses), Stokes simulations (dashed lines), and the thin-film model (solid lines) in the diffusion limit for varying gap height $d$, or no gap at all ($d\to\infty$, unfeasible in the thin-film model). The modelling parameters were adapted to the experiment, their values are noted in the main text. Here, the contact line is pinned in both simulations and experiments but is found to depin in some experimental runs.
    The agreement between Stokes and thin-film simulations is best for small gap heights, as discussed in the context of Fig.~\ref{fig:stokes_volume_height}. Compared to the experiment both simulations underpredict the evaporation time, presumably due to the exclusion of thermal effects in the models.}\label{fig:volume_height_comparison}
\end{figure}

In Fig.~\ref{fig:volume_height_comparison} we depict the normalised volume and height curves as obtained from experiments and simulations for varying values of the gap height $d$. The gap height is color coded, while the line type distinguishes the method (thin-film model / Stokes model / experiment) used for the generation of the data.
It should be noted that the case of having no upper plate at all ($d \to \infty$) is included only for the experiment and Stokes simulations as this limit cannot be achieved in the thin-film model.

As discussed previously, the thin-film and Stokes model agree well, in particular for small gap heights, and their discrepancies are explained by the effect of vertical diffusion in the Stokes model. Further, we find that both models compare well to the experimental results, where again the agreement is best for smaller heights $d$.
Both, in the experiment and in the simulations, the evaporation rate is higher for larger values of the gap height and the drop evaporates most rapidly in the unconfined case $d\to \infty$. This reveals that the experiment is in fact in a diffusion-limited regime, as are the simulations.
The volume is found to decrease linearly over time both in the experiment and in the modelling, as it is expected for the diffusion-limited case.
It is again due to the contact line pinning that the drop height $H$ relates linearly to the volume, i.e., $H(t)$ is also linear. However, this relation relaxes in the experimental data once the drop becomes very shallow and the contact line depins. The depinning causes a deformation of the $H(t)$ curve that resembles exactly our simulations where contact angle hysteresis was included in the thin-film model, namely, Fig.~\ref{fig:thinfilm_pinning}.

Despite the similarities between experiment and theory, the modelling appears to systematically overestimate the evaporation rate even before the depinning occurs. We presume that this is caused by a reduction of the vapour pressure due to cooling by latent heat of evaporation. As both presented models are isothermal, such effects are excluded in the simulations. This is supported by Figure~\ref{fig:Stokes_thermal} in the appendix, where we showcase the same measurements performed with an extended Stokes model that includes heat transfer [model not discussed here, specification follows \citet{Diddens2017} for the case of a simple liquid]. There, most of the discrepancies are explained by the inclusion of thermal effects. The assumption of an isothermal gas phase is more accurate for thinner gap heights, hence the agreement between the isothermal models and experiments (or thermal modelling) is best for small gaps.

\section{Conclusion}
\label{sec:conc}
We have developed an effective two-field thin-film model for sessile shallow drops of volatile liquids under the strong confinement imposed by placement in the narrow gap between two parallel plates. The model consists of coupled time evolution equations for the drop height or film thickness profile and the vertically averaged vapour concentration profile in the narrow gap. A particular strength lies in its ability to describe the coupled liquid convection, liquid-vapour phase change and vapour diffusion for the full spectrum of dynamical coefficients ranging from the diffusion-limited regime at fast phase change to the phase transition-limited regime at slow phase change. In this way it unifies the two groups of isothermal thin-film models available in the literature as reviewed in the introduction.

After presenting the model it has been used to investigate evaporating drops in the different regimes. The obtained numerical results have on the one hand recovered analytical relations we have derived in the two limiting cases. On the other hand they have provided insights into the intermediate regime where the evolution shows aspects of both limiting cases. Furthermore, additionally to the thin-film model we have introduced a model based on the Stokes equation for the liquid in the drop, a diffusion equation for the vapour concentration in the gap and adequate boundary conditions. In particular, the employed modelling of the process of phase transition matches the one in the thin-film model.
Results obtained with the thin-film model and with the Stokes model are compared and, overall, show satisfactory agreement. As expected, the agreement is nearly perfect in all dynamic regimes for narrow gaps and relatively large drops and becomes less so for wide gaps and small drops.

We found that in the long-wave model the dependency on the contact angle can be traced back to a dependency on the liquid surface area. Thereby, the evaporation rate is independent on the contact angle in the diffusion-limited case whereas in the phase transition-limited case the evaporation of drops with smaller contact angle (or larger initial radius $R$) speeds up by $1/R^2$. Note that the former is only true in the constrained gap geometry and not in an unconfined setting. Additionally, the reduced height of smaller contact angle droplets allows for more shallow gap geometries, which improves the agreement of the long-wave results with the full Stokes model.

The good agreement with the Stokes results for narrow gaps for small or moderate contact angles indicates that the developed thin-film model can be used as a valid tool for the study of shallow sessile drops of volatile liquids in a variety of contexts across all dynamical regimes. The model can easily be adapted for many closely related systems, e.g., incorporating chemically or topographically heterogeneous substrates, resulting in contact line pinning and depinning \citep{DeDS2018l}, adding gravity or other body forces \citep{KMDB2020nm}, or considering the importance of the evaporation regime in dip coating \citep{BNFG2017jpcc,DeDG2016epje}. As the presented thin-film description is of gradient dynamics form, necessary changes can often be incorporated via adapting the underlying free energy functional that here only incorporates wetting, interface and bulk energies of the liquid and vapour entropy [cf.~Eq.~\eqref{eq:vapourGap-energy}]. This includes the usage of other equations of state (then adapting the considerations in Section~\ref{sec:tf-real}), employing the full (not long-wave) curvature for the liquid-vapour interface \citep{Thie2018csa}, and the incorporation of more realistic wetting energies \citep{Chur2003cj,HuTA2017jcp}.

Note that the presented thin-film model is isothermal, while the Stokes calculations are presented for both cases, isothermal and non-isothermal. One may incorporate thermal effects into the thin-film model, in the simplest case by using an effective film height-dependent transfer constant as often done for drops on heated substrates in a gas of low thermal conductivity [see discussion of Eq.~(11) in \citep{Thie2014acis}]. Additionally, including thermal Marangoni effects as in \citep{AjGS2010jcis,SaRC2017jfm} and also heat diffusion is challenging as the full dynamics of heat has to be incorporated into the gradient dynamics form. This remains a task for the future.
However, because of its gradient dynamics form, the presented thin-film model may readily serve as a building block in thin-film descriptions of more complex systems. It may, for instance, be combined with recent descriptions of droplets on soft elastic substrates \citep{HeST2021sm,HEHZ2022preprint} and polymer brushes \citep{ThHa2020epjt} to study how the evaporation or condensation of drops is modified by substrate softness and brush properties, respectively. In principle, the approach can also be extended beyond pure liquids to capture evaporation, condensation and absorption of liquid mixtures, e.g., to study the dynamics of mixtures of volatile liquids where selective evaporation results in intriguing effects \citep{ChHS2011prl,KaLR2017l,HKRS2021l}, the evaporation of drops on liquid-infused \citep{GWXM2015l} or porous \citep{Gamb2014cocis} media, or the absorption of binary vapours into polymer brushes \citep{SmEB2020m}.
\\[2em]
\textbf{Acknowledgements.} We acknowledge support by the Deutsche Forschungsgemeinschaft (DFG) via Grants TH781/8-1 and TH781/12. Further, we are grateful to Christopher Henkel for fruitful discussions.\\[2em]
\textbf{Declaration of Interests.} The authors report no conflict of interest.\\[2em]
\textbf{Author ORCIDs.}\\
Simon Hartmann: \url{https://orcid.org/0000-0002-3127-136X};\\
Christian Diddens: \url{https://orcid.org/0000-0003-2395-9911};\\
Maziyar Jalaal: \url{https://orcid.org/0000-0002-5654-8505};\\
Uwe Thiele:  \url{https://orcid.org/0000-0001-7989-9271}.

\bibliographystyle{jfm}
\bibliography{evaporation_gap}

\clearpage
\appendix

\section{Asymptotics of the limit cases}
\label{sec:res-limit}
Instead of using pure numerics, we may also assess some features of the diffusion-limited and transition-limited cases through their asymptotic behaviour. In the following, we derive an estimate for the maximum particle transport rates if either of the limit cases is strongly met. This allows us to obtain the dependency of the effective evaporation rate on the parameters, e.g., the influence of the gap height, and further allows to predict the parameter regime in which we expect the transition between the two limit cases.

For a distinction of the diffusion-limited and the transition-limited dynamics, we consider the particle fluxes associated with the two processes, namely, the diffusive flux $\boldsymbol{j_\mathrm{diff}}$ and the transfer flux $j_\mathrm{evap}$, which can be identified from the dynamical equation for the vapour phase, as given in Eq.~\eqref{eq:film-vapour-evolution-case2simple}:
\begin{align}
  \partial_t[(d-h)\rho] & = - \bnabla\bcdot\boldsymbol{j_\mathrm{diff}} + \rho_\mathrm{liq}\,j_\mathrm{evap}
  = \bnabla\bcdot\left[D (d-h) \nabla\rho\right]
  + \rho_\mathrm{liq}\,j_\mathrm{evap}.\label{eq:vapor-diffusion-flux}
\end{align}
Here, $\rho_\mathrm{liq} j_\mathrm{evap}$ is in units of particles per time per area, while $\boldsymbol{j_\mathrm{diff}}$ is in units of particles per time per cross-section length.

We then obtain the total vapour flux $J_\mathrm{diff}$ away from a macroscopic droplet by integrating over a surrounding closed curve $\partial \Omega$, e.g., a circle of radius $R$ where the film height vanishes:
\begin{align}
  J_\mathrm{diff} & = \oint_{\partial \Omega} \boldsymbol{j_\mathrm{diff}} \, \mathrm{\mathbf{d}}\boldsymbol{s} = -2\upi\,R\,D\,d\,\partial_r \rho\big|_{r=R}\label{eq:total_diffusive_flux}
\end{align}
The vapour density gradient $\partial_r \rho$ may be approximated by solving the diffusion equation in the gap outside of the droplet obtained from Eq.~\eqref{eq:vapor-diffusion-flux}, where we again neglect the adsorption layer, i.e., assume $h\approx 0$ and therefore $j_\mathrm{evap} \approx 0$. Defining the vapour densities at the contact line $\rho_\mathrm{drop}=\rho(r=R)$ and at the far end of the gap $\rho_\mathrm{lab}=\rho(r=L)$, the resulting density profile is
\begin{align}
  \rho(r) & = (\rho_\mathrm{drop} - \rho_\mathrm{lab}) \frac{\log(r/L)}{\log(R/L)} + \rho_\mathrm{lab}.\label{eq:diffusion_eq_solution}
\end{align}
In the case of very rapid diffusion (transfer-limited case), the vapour density near the contact line will be rather close to $\rho_\mathrm{lab}$, whereas for very fast evaporation (diffusion-limited case) it will be close to saturation. Thus, using Eqs.~\eqref{eq:total_diffusive_flux} and~\eqref{eq:diffusion_eq_solution}, we can obtain an upper limit for the total diffusive flux
\begin{align}
  J_\mathrm{diff} & < J_\mathrm{diff}^\mathrm{max} = 2\upi D\,d\,\frac{\rho_\mathrm{sat} - \rho_\mathrm{lab}}{\log(L/R)}\label{eq:max_J_diff}
\end{align}
that also corresponds to the upper limit of the total evaporation rate in the diffusion-limited case. Note that here, due to the constrained geometry of the gap, the diffusion-limited evaporation rate is not proportional to the droplet radius $R$, as found in the case of (unconstrained) spherical diffusion, see e.g., Refs.~\citep{DBDH2000pre,HuLa2002jpcb}, where the linear relation is obtained using a similar argument.

Similarly, we evaluate the total evaporation rate $J_\mathrm{evap}$ by integrating the corresponding terms in Eq.~\eqref{eq:film-vapour-evolution-case2simple} over the domain covered by a macroscopic drop of radius $R$:
\begin{align}
  J_\mathrm{evap} & = \int_{\Omega} \rho_\mathrm{liq}\,j_\mathrm{evap}\,\mathrm{d}^2 r \approx \rho_\mathrm{liq} M \int_{\Omega} \left[ \gamma \Delta h + \rho_\mathrm{liq}k_B T \log \left(\frac{1 - \rho_\mathrm{tot}/\rho}{1 - \rho_\mathrm{tot}/\rho_\mathrm{sat}}\right) \right] \mathrm{d}^2 r,
\end{align}
where in the last step we have neglected the Derjaguin pressure, and used Eq.~\eqref{eq:saturation-condition} to replace $f_\mathrm{liq}$. For a water droplet of \SI{1}{mm} radius of curvature the Laplace pressure (first term in the integral) is of order \SI{72}{Pa}, whereas the order of the second term is governed by $\rho_\mathrm{liq} k_B T \approx \SI{1.4e8}{Pa}$. We conclude that the curvature term may be neglected for any macroscopic drop, i.e., the Kelvin effect only contributes for nanoscopic droplets.

The total transfer rate is maximal when the vapour concentration above the droplet is minimal, i.e., close to the ambient value $\rho_\mathrm{lab}$. In this case, the evaporation occurs rather homogeneously over the entire surface area of the droplet. Hence, for the considered shallow droplets we find the upper limit for the total evaporative flux
\begin{align}
  J_\mathrm{evap} & < J_\mathrm{evap}^\mathrm{max} = \upi R^2 \, \rho_\mathrm{liq}^2 M \, k_B T \, \log \left(\frac{1 - \rho_\mathrm{tot}/\rho_\mathrm{lab}}{1 - \rho_\mathrm{tot}/\rho_\mathrm{sat}}\right).\label{eq:max_J_evap}
\end{align}
Also here $J_\mathrm{diff}$ and $J_\mathrm{evap}$ will be equal. Figure~\ref{fig:thinfilm_volume_height} shows the decrease of the droplet volume over time -- related to the total particle flux by $V(t)=V_0-\int  J_\mathrm{evap}(t)\,\mathrm{d}t$. Hence, if we consider the differential equation
\begin{equation}
  \frac{\mathrm{d} V(t)}{\mathrm{d} t} = J_\mathrm{evap} \sim R(t)^2
\end{equation}
for the phase transition-limited case, we find that the radius $R(t)$ must decrease linearly with time, which is in agreement with the simulations in Fig.~\ref{fig:thinfilm_volume_height}~(c).
Similarly, for the diffusion-limited case, the rate $J_\mathrm{diff}^\mathrm{max} \sim 1 / \log(L/R)$ can be considered independent of the droplet radius $R$ if the lateral extent $L$ of the gap is large compared to $R$. Then, we follow that the drop volume $V(t)$ (not the radius) must decrease linearly with time, as confirmed by Fig.~\ref{fig:thinfilm_volume_height}~(a).

We conclude, that in the diffusion-limited case the time $t_\mathrm{evap}$ until complete evaporation scales with the drop volume $V\sim R^3$. Note that this in particular deviates from the scaling found for spherical diffusion in an unconstrained geometry, where the evaporation time scales as $t_\mathrm{evap}\sim R^2$, which is known as the `$D^2$-law'~\citep{DBDH2000pre, LeVY2014sm, SWVO2016jfm}. In contrast, in the rate-limited case we find that the total evaporation time scales as $t_\mathrm{evap}\sim R$.

In both limit cases, the particle flux will correspond to the upper limits given by Eq.~\eqref{eq:max_J_diff} and Eq.~\eqref{eq:max_J_evap}, respectively.

\begin{figure}
  \centering
  \includegraphics[width=\textwidth]{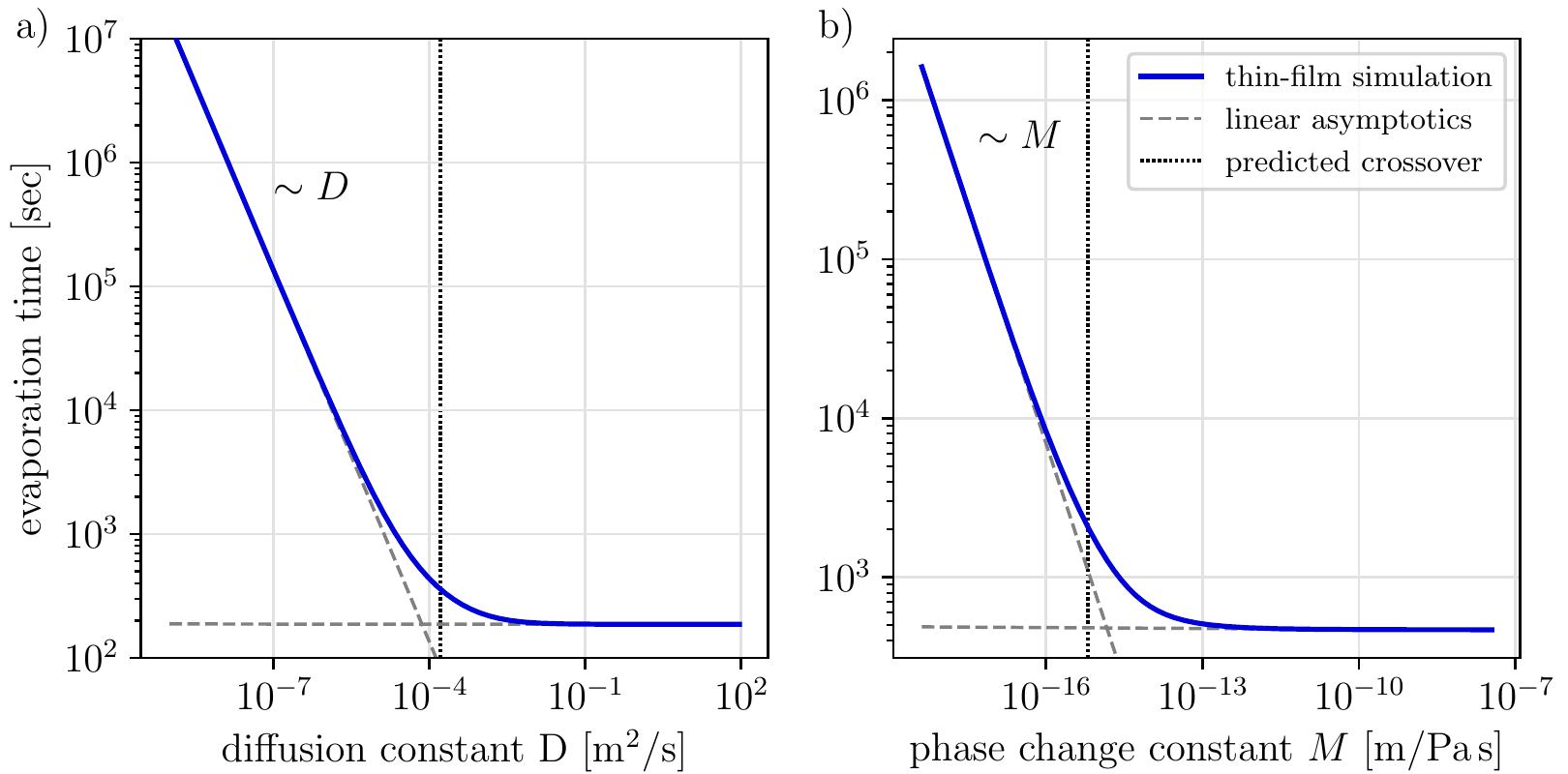}
  \caption{Time until a droplet has completely evaporated as a function of (a) the diffusion constant $D$ and (b) the phase transition rate constant $M$. Each solid blue curve results from 400 simulations of Eqs.~\eqref{eq:film-vapour-evolution-case2simple-polar}. The dashed straight lines indicate the asymptotic behaviour, i.e. constant or linear, revealing either the diffusion-limited or phase transition-limited regime. The vertical lines mark the parameter values at which the crossover between the two regimes is predicted by balancing the theoretical flux rates of diffusion and phase change [Eqs~\eqref{eq:max_J_diff}~\&~\eqref{eq:max_J_evap}]. The values of the fixed parameter are $M=\SI{4e-19}{\meter / \pascal\,\second}$ in (a) and $D=\SI{2.82e-5}{m^2/s}$ in (b).}\label{fig:thinfilm_evap_time}
\end{figure}

In consequence, the crossover between the two regimes will occur when the two values are identical, i.e., at the dimensionless ratio
\begin{equation}
  \frac{d\,D\,(\rho_\mathrm{sat} - \rho_\mathrm{lab})}{R^2\rho_\mathrm{liq}^2 M \, k_B T}
  = \frac{1}{2}   \log \left(\frac{1 - \rho_\mathrm{tot}/\rho_\mathrm{lab}}{1 - \rho_\mathrm{tot}/\rho_\mathrm{sat}}\right)\,\log\left(\frac{L}{R}\right)
  \label{eq:J_cross}
\end{equation}
that only varies logarithmically. The intermediate regime, where diffusion and phase transition act on similar time scales and the evaporation dynamics results from their intricate interplay, is centred about the crossover value. This is illustrated in Fig.~\ref{fig:thinfilm_evap_time}, where we have measured the time required until complete evaporation for a set of 400 simulations differing only in one key parameter, i.e., the diffusion coefficient $D$ in panel (a) or the transfer coefficient $M$ in panel (b).

As expected, we find a transition between the two limiting cases upon variation of either parameter. For low values of the diffusion coefficient $D$, the behaviour is limited by vapour diffusion and therefore depends linearly on $D$. In contrast, for a rapid diffusion the time until complete evaporation is independent of the diffusion constant. Similarly, we find that the evaporation time depends linearly on the phase transition coefficient $M$ if it is small, i.e. limited by phase transition, and is independent on the transition rate if it is orders of magnitude larger. The parameter regime of the transition between the two cases is correctly predicted by Eq.~\eqref{eq:J_cross} and marked in Fig.~\ref{fig:thinfilm_evap_time} with a vertical bar. As a visual guide we also depict dashed lines indicating both linear and constant fits of the relevant parameter to the evaporation time.

\FloatBarrier
\section{Supplementary figures \& tables}

\begin{figure}
  \centering
  \includegraphics[width=.5\textwidth]{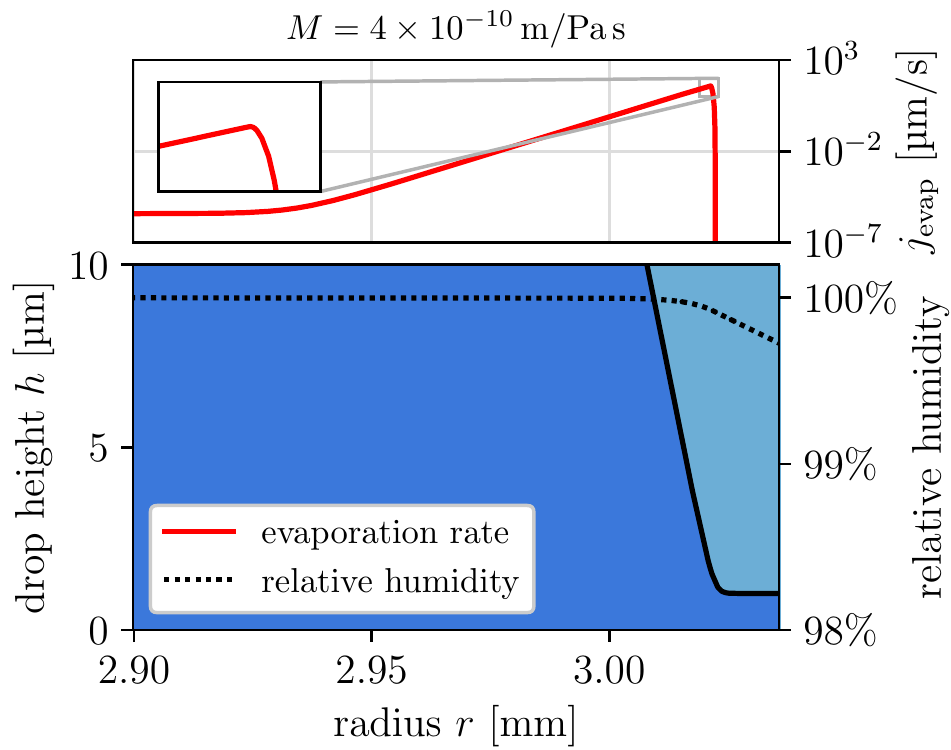}
  \caption{Highly zoomed snapshot of the contact line region of Fig.~\ref{fig:thinfilm_snapshots}~(a) in order both showing the character of the peak in the local evaporation rate $h_\mathrm{evap}$ and the smoothness of the solution in the long-wave model. The apogee of the evaporation rate smoothens over a very short length, as seen in the inset.}\label{fig:thinfilm_snapshot_zoom}
\end{figure}

\begin{figure}
  \centering
  \includegraphics[width=.95\textwidth]{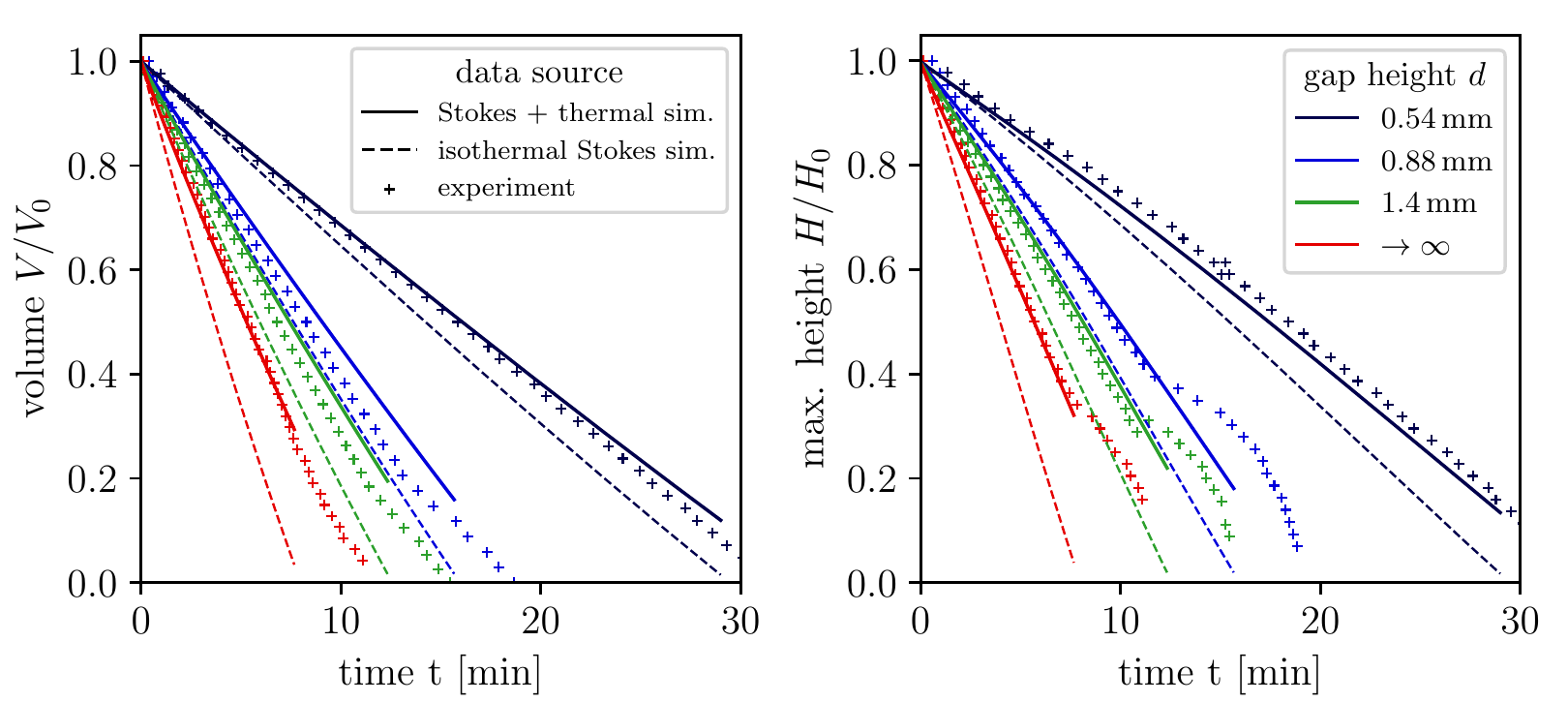}
  \caption{Comparison of the isothermal Stokes model used in this work (dashed lines) and an extended Stokes model that includes heat transfer (solid lines) to the experimental data (crosses), proving that the thermal effects can explain most of the discrepancies between our isothermal models and the experiments.}\label{fig:Stokes_thermal}
\end{figure}

\begin{table}
  \centering
  \begin{tabular}{c c c c c}
    \textbf{gap height $d$~} & \textbf{~drop height~} & \textbf{~drop volume~} & \textbf{~contact angle~} & \textbf{~drop radius} \\
    \SI{0.54}{mm}            & \SI{0.44}{mm}          & \SI{0.42}{\micro l}    & \SI{63.5}{\degree}       & \SI{0.73}{mm}         \\
    \SI{0.88}{mm}            & \SI{0.43}{mm}          & \SI{0.34}{\micro l}    & \SI{66.2}{\degree}       & \SI{0.67}{mm}         \\
    \SI{1.4}{mm}             & \SI{0.45}{mm}          & \SI{0.38}{\micro l}    & \SI{66.1}{\degree}       & \SI{0.69}{mm}         \\
    $\infty$                 & \SI{0.44}{mm}          & \SI{0.47}{\micro l}    & \SI{59.3}{\degree}       & \SI{.79}{mm}
  \end{tabular}
  \caption{Characteristic parameters of the initial drop shape ($t=0$) in the experimental runs for the comparison with simulations in Fig.~\ref{fig:volume_height_comparison}. The values are measured from shadowgraphy images and are then used to initialise the simulations with matching drops.}\label{tab:exp_drop_parameters}
\end{table}

\end{document}